\documentclass[12pt]{article}

\usepackage{cite}
\usepackage{mathrsfs}

\usepackage{epsfig}

\usepackage{amsmath, amsthm, amssymb}

\usepackage{color}

\numberwithin{equation}{section}

\newcommand{\be}{\begin{equation}}
\newcommand{\ee}{\end{equation}}
\newcommand{\beq}{\begin{equation}}
\newcommand{\eeq}{\end{equation}}
\newcommand{\bea}{\begin{eqnarray}}
\newcommand{\eea}{\end{eqnarray}}
\newcommand{\gsim}{\lower.7ex\hbox{$\;\stackrel{\textstyle>}{\sim}\;$}}
\newcommand{\lsim}{\lower.7ex\hbox{$\;\stackrel{\textstyle<}{\sim}\;$}}
\newcommand{\ptwo}{q}
\newcommand{\Pks}{P_s^0}
\newcommand{\kpiv}{k_0}
\newcommand{\phipiv}{\phi_0}
\newcommand{\Ne}{N_e^0}
\newcommand{\A}{A}
\newcommand{\npiv}{n_0}
\newcommand{\kobs}{k_{obs}}
\newcommand{\Mpci}{{\rm Mpc}^{-1}}

\newcommand{\lss}{\Theta_*}
\newcommand{\nl}{\rule{0pt}{18pt}\\}

\newcommand{\run}{{\rm d}n/{{\rm dln} k}\big |_{k_0}}
\newcommand{\runofrun}{{{\rm d^2}n}/{{\rm d ln^2} k}\big |_{k_0}}
\newcommand{\pca}{\varepsilon}
\newcommand{\dchisq}{\Delta \chi^2}

\newcommand{\obs}{d}
\newcommand{\mdl}{\mathcal{M}}
\newcommand{\mlike}{\mathcal{E}}
\newcommand{\NROF}{LOG+NRO$\mathcal{F}$}
\newcommand{\NROG}{LOG+NRO$\mathcal{G}$}

\catcode`@=11
\long\def\@caption#1[#2]#3{\par\addcontentsline{\csname
  ext@#1\endcsname}{#1}{\protect\numberline{\csname
  the#1\endcsname}{\ignorespaces #2}}\begingroup
    \small
    \@parboxrestore
    \@makecaption{\csname fnum@#1\endcsname}{\ignorespaces #3}\par
  \endgroup}
\catcode`@=12

\begin{document}

\vspace{-0.5cm}
\thispagestyle{empty}
\begin{flushright}
{\small IFT-UAM/CSIC-07-60}\\
{\small CERN-PH-TH/2007-225}
 \end{flushright}

\vspace*{1.0cm}

\begin{center}
{\Large\textbf{Flat Tree-level Inflationary Potentials\\
in Light of CMB and LSS Data}} \vspace*{0.8cm}

\textbf{G. Ballesteros}  $^{{\rm a},{\rm b},}
$\footnote{guillermo.ballesteros@uam.es,
$^2$alberto.casas@uam.es,
$^3$jose.espinosa@cern.ch,\hspace*{2cm}\\
\hspace*{0.37cm} $^4$rruiz@delta.ft.uam.es,
$^5$rxt@astro.ox.ac.uk},
\textbf{J. A. Casas} $^{{\rm a},2}$
\textbf{J. R. Espinosa} $^{{\rm a},3}$,\\
\textbf{R. Ruiz de Austri} $^{{\rm a},{\rm c},4}$ 
and \textbf{R. Trotta} $^{{\rm d},5} $

\vspace{0.5cm}
$^{\rm a}$ {\it IFT-UAM/CSIC, UAM,
Cantoblanco, 28049 Madrid, Spain}

\vspace{0.2cm}
$^{\rm b}$ \textit{CERN, Theory Division, CH-1211 Geneva 23, Switzerland}

\vspace{0.2cm}
$^{\rm c}$ \textit{Dept. de F\'isica Te\'orica, UAM, Cantoblanco, 28049
Madrid, Spain}

\vspace{0.2cm} $^{\rm d}$ \textit{Oxford University, Astrophysics
Department,\\ Denys Wilkinson Building, Keble Road, Oxford OX1
3RH, UK}

\end{center}

\vspace{0.5cm}
\begin{center}
{\bf Abstract}
\end{center}

We use cosmic microwave background and large scale structure data to test a 
broad and physically well-motivated class of inflationary models: those with 
flat tree-level potentials (typical in supersymmetry). The non-trivial 
features of the potential arise from radiative corrections which give a 
simple logarithmic dependence on the inflaton field, making the models very 
predictive. We also consider a modified scenario with new physics beyond a 
certain high-energy cut-off showing up as non-renormalizable operators (NRO) 
in the inflaton field. We find that both kinds of models fit remarkably well 
CMB and LSS data, with very few free parameters. Besides, a large part of 
these models naturally predict a reasonable number of e--folds. A robust 
feature of these scenarios is the smallness of tensor perturbations ($r 
\lsim 10^{-3}$). The NRO case can give a sizeable running of the spectral 
index while achieving a sufficient number of e--folds. We use Bayesian model 
comparison tools to assess the relative performance of the models. We 
believe that these scenarios can be considered as a standard physical class 
of inflationary models, on a similar footing with monomial potentials.

\newpage
\setcounter{page}{1}

\section{Introduction}

The WMAP 3--years data \cite{Peiris:2003ff,Spergel:2006hy} on the
cosmic microwave background radiation (CMB) represent a milestone
in the history of observational cosmology due to their great
precision. The challenge for theoretical cosmology and particle
physics is to understand these data within some theoretical
framework and to make predictions to be tested by future
observations. In this respect, inflation stands as the most
successful and promising theoretical scenario. Proposed in 1981 as
a solution to the flatness and horizon problems~\cite{Guth:1980zm}
it was later shown to predict an almost scale invariant spectrum
of matter perturbations~\cite{Mukhanov:1981xt, Starobinsky:1982ee,
Hawking:1982cz, Guth:1982ec, Bardeen:1983qw}, making it a very
successful idea. However, we still lack a concrete description of
the mechanism producing inflation in the very early universe. The
simplest implementation postulates a real scalar field, $\phi$,
whose homogeneous part drives inflation during its way towards the
minimum of its potential, $V(\phi)$. Therefore, to provide a
description of the inflationary epoch of the universe we need to
elucidate the functional form of the inflaton potential
\cite{Lidsey:1995np}. In this respect it is fair to say that,
despite interesting developments, there is still no convincing
inflationary scheme based on particle physics.

The main goal of this work is to compare a broad class of models,
which are physically well motivated, with CMB and large scale
structure (LSS) data. As shown below, these models predict a
scalar spectral index with non-constant slope, in contrast with
the ans\"atze used in standard analyses. Interestingly, these
models are very predictive: the number of independent parameters
is comparable to the free parameters used in more
phenomenological approaches.

This paper is organized as follows. In section 2 we review the
standard way of comparing inflationary potentials with
cosmological data through parametrizations of the primordial
scalar and tensor power spectra that involve the spectral indices
and their derivatives (up to a given order). In section 3 we
introduce the class of models we are interested in, namely,
models with flat tree-level potentials which are generically
lifted in a controlled way by radiative corrections. This type of
models is then extended by taking into account possible effects
from non-renormalizable operators that appear after integrating
out some heavy new physics. In both cases we discuss the physical
motivation and significance of such models. Section 4 is devoted
to the derivation of  the power spectra for the two classes of models
as a function of suitable model parameters. The data analysis
procedure is discussed in section 5; we present parameter
constraints within each class of models from our analysis in section
\ref{results} while section 7 is devoted to the issue of model
comparison, assessing which model is in better agreement with the
data. Finally, our conclusions are presented in section 8. In
addition, the appendix contains some technical details relevant to
the analysis of the model with non-renormalizable operators.

\section{Comparing inflationary potentials with data}

The usual strategy to probe inflationary potentials is the
comparison of theoretical predictions for the spectrum of
primordial (scalar and tensor) cosmological perturbations
[$P_s(k)$ and $P_t(k)$ respectively, where $k$ stands for the
space wave-number] with CMB and LSS data. This is usually done in
the slow-roll approximation \cite{Liddle:1994dx}. Generically, in
this situation $\left|\epsilon, \eta, \xi\right| \ll 1$, with \bea
\label{slowrollpar} \epsilon\equiv\frac{1}{2}M_p^2
\left(\frac{V'}{V} \right)^2\ ,\;\;\; \eta\equiv M_p^2
\frac{V''}{V} \ ,\;\;\; \xi\equiv M_p^4 \frac{V'V'''}{V^2} \;,
\eea where primes denote derivatives with respect to the inflaton
$\phi$ and $M_p\equiv m_p/\sqrt{8\pi}$ (being $m_p\simeq 10^{19}$
GeV) is the reduced Planck mass. In this approximation \be
\label{Pk1} P_s(k)\simeq
\frac{1}{24\pi^2\epsilon}\frac{V}{M_p^4}\,, \ee \bea \label{Pt}
P_t(k)\simeq \frac{3}{2\pi^2}\frac{V}{M_p^4}\ . \eea Thus the
tensor to scalar ratio simply reads \be \label{tensc}
r\equiv\frac{P_t}{P_s}\simeq 16\epsilon\,. \ee The relationship
between the inverse distance scale $k$ and the inflaton field
$\phi$ is given in slow-roll by
\bea
\label{dlnk}
\frac{d \phi}{d \ln k}\simeq -M_p\sqrt{2\epsilon}\ .
\eea
The number of e-folds between the beginning of inflation ($t_*$)
and the end ($t_e$) is
\be
\label{efoldsdef}
N_e = \int_{t_*}^{t_e} {H} d t \simeq \frac{1}{M_{p}^2}\int_{\phi_{e}}
^{\phi_*} \frac{V}{V'} d\phi = \frac{1}{M_{p}}\int_{\phi_{e}}
^{\phi_*} \frac{1}{\sqrt{2\epsilon}} d\phi \, ,
\ee
where we have used the slow-roll approximation to change variables
from $t$ to $\phi$. The usual requirement to solve the horizon problem is $N_e\simeq 50-60$ although the precise number
depends on the details of the post-inflationary epoch
\cite{Lyth:1998xn}. Of course, the total number of e-folds may be
much larger, but only 50-60 e-folds are needed and actually WMAP
only probes directly the first $\sim 7$ e-folds, which can be
extended to $\sim 10$ by using LSS data and other observations.
Thus the subscript '$*$' above denotes the time at 50-60 e-folds
before the end of inflation. On the other hand, there might be
several episodes of inflation, in order to achieve the required
50-60 e-folds. In that case, WMAP would be only sensitive to the
first(s) of these episodes after $t_*$.

An important quantity when performing fits to the observations is
the scalar spectral index, $n$, which describes the variation of
$P_s$ with $k$,
\bea
\label{n}
n - 1 = \frac{d \ln P_s}{d \ln k}\simeq 2\eta-6\epsilon\ .
\eea
The spectral index itself may change with $k$:
\be
\label{dndlnk}
\frac{d n}{d\ln k}\simeq -2\xi +16\epsilon\eta-24\epsilon^2\ .
\ee
The right-hand sides of (\ref{n}) and (\ref{dndlnk}) correspond to
the slow-roll approximation. Similar equations, involving the
tensor spectral index $n_t(k)$, can be written for the variation
of $P_t(k)$ with $k$. Summarizing, observational information about $P_s$
(and thus about $n$ and ${d n /d\ln k}$) gives information about
$V$ and its derivatives, $V'$, $V''$, $V'''$ and so on, providing
a link between theory and experiment.

\vspace{0.4cm} \noindent {\bf Standard Parametrizations of the
Power Spectrum}

\vspace{0.3cm} \noindent

A common approach to determining the power spectrum from CMB and
LSS data is to perform a Taylor expansion of $\ln P_s(k)$ and $\ln
P_t(k)$ in $\ln k/k_0$ around zero, where $k_0$ is a pivotal scale (from now
on fixed to $k_0\equiv 0.002\ {\rm Mpc}^{-1}$), \bea \label{stp}
\ln P_s(k)&=& \ln P_s(\kpiv)+
  \left[n(k_0)-1\right]\ln\frac{k}{\kpiv}+\frac{1}{2}\frac{dn}{d\ln
  k}{\bigg|}_{\kpiv}\left(\ln\frac{k}{\kpiv}\right)^2 + \cdots\,, \\
\label{stp2}
\ln P_t(k)&=& \ln P_t(\kpiv)+
  n_t(k_0)\ln\frac{k}{\kpiv} + \cdots\,.
\eea Note from equations~(\ref{Pk1}) and (\ref{tensc}) it follows
that, at first order in slow-roll \cite{Copeland:1993jj}: \be
r(k_0)\equiv\frac{P_t(\kpiv)}{P_s(\kpiv)}\simeq-8n_t(k_0)\ . \ee
Usually, one does not go beyond the order shown in the
equations~(\ref{stp}) and (\ref{stp2}). In particular, a running
of the tensor spectral index has not been considered because
presently the tensor contribution to the spectrum is only weakly
constrained by the data. Therefore one typically fits four independent
parameters, namely: $\{\ln P_s, n, d n/d \ln k, r\}_{k_0}$ and often
the running of the spectral index, $d n/d \ln k|_{k_0}$, and the
tensor to scalar ratio, $r(k_0)$, are set to zero. The improvement in
the fit obtained when introducing the latter parameters is neither
large enough to be considered as a strong indication for their
presence nor small enough to be considered as irrelevant. In
general, the issue of whether an extra parameter is needed or not
is a difficult one and has to be addressed with care, see
\cite{Gordon:2007xm,Trotta:2005ar} for a discussion. We return to
this question in Section~\ref{sec:modcomp}.

The Taylor expansion above has been used by the WMAP
collaboration. Assuming a $\Lambda$CDM universe and setting $d n/d
\ln k = r = 0$, WMAP 3--yr data alone give
\cite{legacyweb}
\be
\label{n1}
n=0.958\pm 0.016\ ,\;\;\;\;(68\%\ {\rm c.l.})
\ee
Including a constant $d n/d \ln k$ the result of the analysis is
\bea
\label{n2}
n(k_0)&=&1.050^{+0.059}_{-0.058}\ ,\;\;\;\;(68\%\ {\rm c.l.})\\
\frac{d n}{d \ln k} &=&  
-0.055^{+0.030}_{-0.031}\;\;\;\;(68\%\ {\rm c.l.})\,.
\label{n3} \eea
The absolute magnitude of $P_s$ depends slightly on the
inflationary model, but roughly one finds $P_s(k_0)\simeq 2\times
10^{-9}$. The WMAP collaboration used these results to probe
monomial potentials $V(\phi)\propto \phi^\alpha$  (with
$\alpha=2,4$) \cite{Spergel:2006hy}. These models predict a
negligible running of $n$, so they are well approximated by
equations~(\ref{stp}) and (\ref{stp2}) with  ${d n / d \ln k} =0$.
The fits seem to exclude $\alpha=4$ \cite{Hamann:2006pf} and any other 
higher monomial
power. The quadratic case, $V=\frac{1}{2} m^2\phi^2$, works quite
well, although it requires very large values of the inflaton
field, $\phi\sim 14\ M_p$. For the purposes of our discussion it
must be stressed that the simple functional forms assumed
previously for $P_s(k)$ and $P_t(k)$ may not be accurate enough to
describe the actual power spectrum of other inflaton potentials
which are well motivated physically. In this sense, although
(\ref{stp}) and (\ref{stp2}) can be useful as phenomenological
approximations, it is important to be open to other
parametrizations. In this paper we will show explicit examples of
this. At the end of the day, the best fit together with the best
physical motivation will determine the preferred functional form.

\section{A broad class of models: Flat Tree-Level Potentials}

We will consider models that have ``{\em flat tree-level
potentials}"\;\footnote{ For a more complete discussion of the
theoretical aspects of these models see \cite{Ballesteros:2005eg}.},
i.e. \be V_{\rm tree}(\phi)=\rho_{\rm tree}={\rm constant}\ . \ee Then, the
potential derivatives $V'$, $V''$,... arise from the
radiative corrections to $V$. These potentials appear typically in
supersymmetric (SUSY) theories: $V_{\rm tree}^{\rm SUSY}$
ordinarily has plenty of {\em accidental} flat directions. A
familiar example of this is the minimal supersymmetric standard
model (MSSM). Such
accidental flatness is broken by radiative corrections since there
is no symmetry protecting it. Generically, at one-loop
\bea
\label{Vgeneric}
V(\phi) = \rho + \beta\ \ln\frac{m(\phi)}{Q}\ ,
\eea
where $Q$ is the renormalization scale (which might have absorbed finite
pieces) and $m(\phi)$ is the most relevant
$\phi$-dependent mass in the spectrum. Note that $\rho$ depends
implicitly on $Q$ through its renormalization group equation (RGE) and
that the $Q$-invariance of the effective potential implies
\bea
\label{betarho}
\beta =\frac{d\rho}{d \ln Q}\ ,
\eea
at one-loop. From now on we will assume $\beta>0$, which is the usual
situation, though the opposite case is also possible and the analysis is
similar.

Since these ``almost flat" directions are so common in SUSY
scenarios, they are natural candidates to drive inflation,
provided the potential stores large enough energy density. As a
matter of fact, particular examples of these approximate flat
directions have been used in the literature to implement
inflation, e.g. the first D-hybrid inflation model belongs to this
class \cite{Binetruy:1996xj}. The important point is that, whatever the
model considered, the slope of $V(\phi)$, and thus the dynamics of
inflation, are determined by radiative corrections. Since the
latter have a very generic functional form (logarithmic), it is
possible to make very model-independent predictions without
relying on a particular model \cite{Ballesteros:2005eg}. Next we work out
these
statements in a more detailed way.

The leading-log approximation (which amounts to summing up the leading-log
contributions to all loops) is implemented in this context
by simply taking $Q=m(\phi)$. This choice eliminates the potentially large
(and thus dangerous) logarithms, improving the convergence of the
perturbative expansion.
Then
\be
\label{Vphi}
V(\phi) \simeq \rho[Q=m(\phi)]  \ .
\ee
In general, one expects
$m^2(\phi) = M^2 + c^2 \phi^2$, where $M$ does not depend on $\phi$,
and  $c$ is some coupling constant (which depends on $Q$ according to
its own RGE). Normally one considers a range of $\phi-$values
where either the constant $M^2$
piece (if it exists) or the $\phi-$dependent part dominate. We assume
we are in the second case. Hence, we will ignore the
possible presence of $M$ and take $Q=m(\phi)=c\,\phi$.

\vspace{0.4cm}
\noindent
{\bf Logarithmic regime (LOG)}

\vspace{0.3cm}
\noindent
In the regime of very small coupling constants one has
$d \beta/d \ln Q\ll \beta$ since the former is higher
order in the couplings and has a loop suppression factor.
Then we can consider $\beta$ as
constant in the range of $Q\propto \phi$ of interest (which
is never too wide). Now
the scalar potential (\ref{Vphi}) can be easily written in
terms of its value at $\phi_0$, the value of the inflaton at $k_0$, integrating equation~(\ref{betarho}):
\bea
\label{VSCR}
V(\phi) = \rho_0 + \beta\ \ln\frac{\phi}{\phi_0}\ ,
\eea
where $\rho_0\equiv \rho(\phi_0)$. Note that (\ref{VSCR}) can also be obtained
from (\ref{Vgeneric}) by choosing $Q=c\phi_0$.

Since $\ln \phi = \lim_{\alpha\rightarrow 0}\alpha^{-1}(\phi^\alpha-1)$
the logarithmic potential (\ref{VSCR}) can be considered in many respects
as a monomial potential
with $\alpha=0$. In particular all the derivatives, which are crucial
for the cosmology of the model, follow that pattern. On the other hand,
as argued above, this potential is physically as well motivated as the
monomial forms.

\vspace{0.4cm}
\noindent
{\bf Logarithmic regime + non-renormalizable operator (LOG+NRO)}

\vspace{0.3cm}
\noindent
If there is a scale of new physics, $M$, higher than the scales
relevant to inflation (i.e. $M^2\gg\phi^2$), the new physics
will generically show up in the effective theory at lower scales
as non-renormalizable operators (NROs) of the light fields, suppressed by
inverse
powers of $M$.
Due to the suppression factor, the impact of the NROs in the physics
at low scales is usually very small. However, if the NRO has
characteristics not shared by the low-energy physics, its effect
may be significant (as happens with higher-dimension operators that
mediate proton decay or give a Majorana mass to neutrinos).
In our case, the new physics does not need to respect the
accidental flat directions of the effective theory. Thus one expects
the inflaton potential (\ref{VSCR}) to become
\be
V(\phi)=\rho_0+\beta
\ln\frac{\phi}{\phi_0}+\phi^4\frac{\phi^{2N}}{M^{2N}}\;.
\label{VNRO}
\ee
The first two terms correspond to the generic one-loop
potential in the small-coupling regime while the last term is a
non-renormalizable operator left in the low-energy theory after
integrating out some unspecified physics at the high scale $M$.
Notice that this scale absorbs any possible coupling in front of the
operator. Of course $V(\phi)$ may contain other NROs of different
order. Here we assume that the one shown in equation~(\ref{VNRO}) is
the lowest order one, and thus the dominant one. The sign and
power we have assumed for this NRO guarantee the
stability of the potential. Notice also that an even power for
this operator is what one expects generically in supersymmetric
theories. An explicit example of this is given in
\cite{Ballesteros:2005eg}. Apart from that, the potential
(\ref{VNRO}) is completely general and therefore the analysis is
quite model-independent.

\section{Primordial spectra in the slow-roll approximation}

\subsection{Logarithmic regime (LOG)}
\label{sec:paramsSCR}

From $V(\phi)$, as given in equation~(\ref{VSCR}), the first three
slow-roll parameters (\ref{slowrollpar}) in the LOG scenario are
\be \label{epsiloneta2} \epsilon\simeq
\frac{1}{2}\ptwo^2\frac{M_p^2}{\phi^2}  \;,\;\;\;\;\;\; \eta \simeq
-\ptwo\frac{M_p^2}{\phi^2}\simeq -2 \frac{\epsilon}{q} \;,\;\;\;\;\;\;
\xi\simeq 2 \eta^2 \,, \ee where we have introduced the quantity
 \be \label{defq}
 \ptwo\equiv\frac{\beta}{\rho_0}\ .
 \ee
Note that $q>0$ since we
have assumed $\beta>0$. Satisfying the
slow-roll condition $\epsilon\ll 1$,
requires $q\ll 1$
 for $\phi < M_p$. This implies in turn $\epsilon\ll |\eta|$.
We give a more quantitative evaluation of this
hierarchy in Section~5.

The number of e-folds between time $t_0$ (i.e.
the time when $\phi=\phi_0$) and the end of inflation, $t_e$,
can be easily computed using
$\epsilon$, as given by equation~(\ref{epsiloneta2}), in the usual
expression (\ref{efoldsdef}):
\bea \label{N0} N_e(t_0\rightarrow t_e)\simeq
-\frac{1}{2}\left[\frac{1}{\eta(\phi_0)}-\frac{1}{\eta(\phi_{e})}\right]
\simeq -\frac{1}{2\eta(\phi_0)} \simeq \frac{1}{2} \frac{\phi_0^2}{q
M_p^2}\equiv N_e^0\ , \eea
where we have used the fact
that inflation comes to an end when $\eta\sim {\cal O}(1)$ (it
could end before that time if inflation is interrupted by other
mechanisms, like a waterfall field in hybrid models).
The $N_e^0$ parameter defined above, besides giving an excellent approximation
to $N_e(t_0\rightarrow t_e)$, will play a relevant role when
performing the fits to the data.

Now the $\phi-k$ connection, equation~(\ref{dlnk}), can be
integrated at first order in $q$, giving \be \label{phik} \phi^2 =
\phi_0^2\left(1 - \frac{1}{N_e^0}\ln \frac{k}{k_0}\right)\ , \ee
where we have used (\ref{N0}). Note that increasing $\phi$
corresponds to decreasing $k$ so that the scales probed by WMAP
correspond to the highest values of $\phi$ during its slow-roll towards
the origin. Now we can straightforwardly
evaluate $P_s(k)$ from equation~(\ref{Pk1}). For the purpose of
comparing the model with the data, it is convenient to write $P_s$
in terms of $P_s^0\equiv P_s(k_0)$ using the general expression
\be
\label{PsPs0}
\ln P_s= \ln P_s^0 + 3\ln\frac{V(\phi)}{V(\phi_0)}
-2\ln\frac{V'(\phi)}{V'(\phi_0)} \ .
\ee
Using equations~(\ref{VSCR}) and (\ref{phik})
and expanding in $q$ we find, at first order, \bea \label{PsSCR}
\ln P_s(k)= \ln P_s^0 + \left(1+\frac{3}{2}q \right) \ln\left(1-
\frac{1}{N_e^0}\ln\frac{k}{k_0} \right)\ , \eea where, from
equation~(\ref{Pk1}), \be \label{Ps0}
P_s^0=\frac{1}{12\pi^2}\frac{\rho_0^3\phi_0^2}{M_p^6\beta^2}\ . \ee The
same result can be obtained by integrating the slow-roll equation
(\ref{n}).

Similarly, the spectrum of tensor perturbations, $P_t(k)$, can be
obtained from equation~(\ref{Pt}) \bea \label{PtSCR} P_t(k)\simeq
\frac{4 q
P_s^0}{N_e^0}\left[1+
\frac{\ptwo}{2}\ln\left(1-\frac{1}{\Ne}\ln\frac{k}{\kpiv}\right)\right]\,.
\eea At the same level of approximation, the tensor-to-scalar ratio
(\ref{tensc}) reads \be \label{simpexp}
r(k)\simeq\frac{4\ptwo}{\Ne}
\left(1-\frac{1}{\Ne}\ln\frac{k}{\kpiv}\right)^{-1}\,.
\ee

Let us now count the number of independent parameters. The power
spectra, $P_s(k)$ and $P_t(k)$, contain three independent
parameters, \{$P_s^0$, $q$, $N_e^0$\}, which are combinations of
the initial parameters \{$\phi_0$, $\rho_0$, $\beta$\}.
Incidentally note that the scalar potential (\ref{VSCR})  is a
function of just two combinations of parameters, but a third one
appears in the conversion of $\phi$ into $k$ through
equation~(\ref{phik}). Actually, the $q$ term in (\ref{PsSCR}) is
subdominant because $q\ll 1$. Removing it from the expression
is a good approximation and eliminates one parameter. So the
expression of $P_s(k)$ contains basically two parameters. This is
to be compared with the three parameters (two if the running of
$n$ is set to zero) of the simple standard parametrization
(\ref{stp}). As a consequence this LOG scenario is highly predictive.
On the other hand, the fits to WMAP data prefer $P_t(k)\ll
P_s(k)$, which means that $P_t$ turns out to be scarcely important
in the fit, and so the number of relevant parameters continues to
be two. Indeed, from equation~(\ref{simpexp}) and $q\ll 1$ we do
expect by construction $P_t(k)\ll P_s(k)$, something that cannot
be postulated from the simple parametrizations (\ref{stp}) and
(\ref{stp2}), which contain the additional parameter $P_t(k_0)$ or,
equivalently, $r$ (unless $r$ is set to zero by hand).

Equations~(\ref{PsSCR}) and (\ref{PtSCR}) summarize the predictions
of models with flat tree-level potentials in the small coupling
regime. It is possible to gain some intuition about the performance of
these functional forms as follows. As mentioned above,
$\epsilon\ll|\eta|$, which means $n-1\simeq 2\eta$. Therefore
[see equation~(\ref{epsiloneta2})],
\bea
\label{smallrunning}
\frac{d n}{d \ln k}\simeq-2\xi = -(n-1)^2\ .
\eea
As a consequence, ${d n/d \ln k}$ is negative, as suggested by the
data, though its value tends to be quite small. In fact the
sign of $n-1$ cannot change along the inflationary process in this class of models. Since the
dependence
of $n$ on the scale is weak, and therefore $n\simeq$ constant, we can
expect a fit similar to the one obtained by using the standard
parametrization of equation~(\ref{stp}) with $dn/d\ln k =0$,
leading to $n_0\sim 0.95$. We will
see in Section~5 that this is indeed the case.

Equation~(\ref{smallrunning}) can be written
in an integrated form as
\bea
\label{smallrunning2}
n =1 -\frac{1}{N_e^0 - \ln (k/ k_0)}
\ ,
\eea
where we have used equation~(\ref{N0}). Equation~(\ref{smallrunning2}) is
the prediction for the spectral index in scenarios with flat tree-level
potential in the regime of small coupling. It can be compared with
the $n=$ constant or $dn/d\ln k=$ constant assumptions
made in standard analyses. Note that $N_e^0$ is the only independent
parameter in (\ref{smallrunning2}) and has a precise physical
meaning.

\subsection{Logarithmic regime and non--renormalizable operator \\ (LOG +
NRO)}

In this scenario the derivatives of the
potential $V(\phi)$ [equation~(\ref{VNRO})] with respect to $\phi$ read
\bea V'(\phi)&=&\frac{\beta}{\phi}+2(N+2)\phi^3\frac{\phi^{2N}}{M^{2N}}\
,\nonumber\\
V''(\phi)&=&-\frac{\beta}{\phi^2}+2(N+2)(2N+3)\phi^2\frac{\phi^{2N}}{M^{2N}}\
,\nonumber\\
V'''(\phi)&=&2\frac{\beta}{\phi^3}+4(N+2)(2N+3)(N+1)\phi\frac{\phi^{2N}}{M^{2N}}\
. \label{derivatives} \eea
The corresponding expressions for $\epsilon, \eta$ are
\bea
\epsilon &=& \frac{1}{2} \frac{M_p^2}{\phi^2}\left[
\frac{\beta}{\rho_0}+
2(N+2)\frac{\phi^4}{\rho_0}\frac{\phi^{2N}}{M^{2N}}\right]^2
\ ,\nonumber\\
\eta &=& \frac{M_p^2}{\phi^2}\left[
-\frac{\beta}{\rho_0}+2(N+2)(2N+3)\frac{\phi^4}{\rho_0}
\frac{\phi^{2N}}{M^{2N}}\right]
\ . \label{SR_NRO}
\eea
In consequence the NRO can have a
significant impact on inflation when the small number
$(\phi/M)^{2N}$ is comparable in size to $\beta/\phi^4$.
It is also immediate to realize from equations
(\ref{derivatives}, \ref{SR_NRO}) that, for sufficiently large $\phi$, the
higher derivatives $V'', V'''$ (and thus $\eta, \xi$) can receive
a large contribution  from the NRO while the contribution to $V'$
(and thus $\epsilon$) is much less significant, thanks to the
additional $(2N+3)$ and $(2N+3)(2N+2)$ factors in $V'', V'''$.
Therefore, one expects modifications of the spectral index, $n$, and
its running $dn/d\ln k$ (which depend on $\eta, \xi$), especially
at the initial scales (very small $k$ and thus large $\phi$), and
much smaller changes in $\epsilon$ (and hence in $N_e$).

Let us now calculate the expression of the power spectrum $P_s(k)$
in this scenario. As in the previous subsection, we start with the
general expression (\ref{PsPs0}) where $V(\phi)$ and $V'(\phi)$ are given now by
equations (\ref{VNRO}) and (\ref{derivatives}) respectively, and
\be
P_s^0=\frac{1}{12
\pi^2}\frac{[V(\phi_0)]^3}{M_p^2[V'(\phi_0)]^2}\  . \ee
Next we
have to convert $\phi$ into $k$ by integrating
equation~(\ref{dlnk}). This is done in the Appendix in an exact
way. The corresponding formula, equation~(\ref{long}), is rather
cumbersome. However, using $|q|\ll 1$, one can write the much simpler but 
extremely accurate expression
\begin{equation}
\label{aproxx1}
\ln\frac{k}{\kpiv}\simeq
-\varphi\Ne\;\,{_2F_1}\left[\frac{1}{N+2},1;
\frac{N+3}{N+2};-\left(A\varphi\right)^{N+2}\right]
{\Bigg|}^{\phi^2/\phipiv^2}_{\varphi=1}\ ,
\end{equation}
where $_2F_1(a,b;c;z)$ is the Gauss' hypergeometric function
\cite{handbook}, $\varphi$ is just a dummy variable and 
\be
\label{A} 
\A^{N+2}\equiv 2(N+2) \frac{\phipiv^{4}}{\beta}
\frac{\phipiv^{2N}}{M^{2N}} \ . 
\ee 
Note that $N_e^0$ is still defined as in equation \eqref{N0} and gives a 
good estimate of the number of e-folds $N_e(t_0\rightarrow t_e)$. It is 
possible to invert (\ref{aproxx1}) numerically to get $\phi=\phi(k)$. 
Plugging $\phi(k)$ into equation~(\ref{PsPs0}) we obtain $P_s(k)$. This is 
the procedure we have followed in doing the fits. Using $\phi(k)$ we can 
also obtain other quantities of interest as functions of $k$, e.g. the 
spectral index $n \simeq 1+2\eta-6\epsilon$ or the tensor to scalar ratio 
$r\simeq 16\epsilon$.

In order to get some intuition about the shapes of $P_s(k)$ and
$n(k)$ it is convenient to derive an analytical approximation to
the previous numerical procedure. Actually the numerical part just
comes from the $\phi$ to $k$ conversion, i.e. the integration of
equation~(\ref{dlnk}). This equation depends on $\epsilon$ (on $V$ and $V'$) but not on higher derivatives of $V$, which are
the ones most affected by the presence of the NRO, as
discussed after equation~(\ref{derivatives}). Therefore it is
sensible to use here the value of $\epsilon$ when the NRO is
switched off, i.e. that of equation~(\ref{epsiloneta2}). Then the
$\phi$ to $k$ relation is still given by equation~(\ref{phik}).
Substituting this in the general expression for $P_s(\phi)$,
equation~(\ref{PsPs0}), and expanding at first order in the NRO
contributions, one gets\footnote{
This approximate formula gives $P_s(k)$ with a maximum error of $\lsim 
13$\% in the most extreme cases although typically is much better. Anyway, 
we remark that in the fit we evaluate $P_s(k)$ numerically using 
equation (\ref{aproxx1}).}
\begin{eqnarray}
\label{appr22}
\ln {\frac{P_s(k)}{\Pks} }
\simeq\left(1+\frac{3}{2}{\ptwo}\right)\ln\left(1-\frac{1}{\Ne}
\ln\frac{k}{\kpiv}\right)
+ \frac{\gamma N_e^0}{N+2}\left[1-\left(1-\frac{1}{\Ne}\ln
\frac{k}{\kpiv}\right)^{N+2}\right]\,,
\end{eqnarray}
where
\be
\label{gamma}
\gamma \equiv
\left\{2N+3+\left[
\frac{1}{2(N+2)}-3\right]\ptwo+
\frac{3}{2(N+2)}{\ptwo^2}\right\}\frac{A^{N+2}}{\Ne}
\approx (2N+3)\frac{A^{N+2}}{\Ne}\,.
\ee
The approximate equality
in the last expression is justified by the smallness of
$\ptwo$. Alternatively, one can expand $n-1 \simeq
2\eta-6\epsilon$ at first order in the NRO and approximate again
the $\phi$ to $k$ conversion by equation~(\ref{phik}). One obtains
\begin{equation}
\label{nbase}
n(k)-1 \simeq
-\left(1+\frac{3}{2}{\ptwo}\right)\frac{1}{\Ne}
\left(1-\frac{1}{\Ne}\ln\frac{k}{\kpiv}\right)^{-1}+
\gamma\left(1-\frac{1}{\Ne}\ln\frac{k}{\kpiv}\right)^{N+1}\,.
\end{equation}
Then the direct integration of (\ref{nbase}) gives back expression
(\ref{appr22}). Equation~(\ref{nbase})  corresponds to a running $n(k)$
with
{\em non}-constant slope, departing from the assumption of
analyses done using the standard parametrization. Unlike in the
LOG scenario, in this case the running is not constrained to be
very small.

It is also worth mentioning that, due to the positivity of
$1-(1/\Ne)\ln{k}/{\kpiv}$, the sign of the LOG and the NRO
contributions to $\{n(k)-1,\ dn/d\ln k,\ d^2n/d\ln^2 k \}$ are
$\{-,\ -,\ -\}$ and $\{+,\ -,\ +\}$ respectively [see
equation(\ref{nbase})]. Since a sizeable running at low $k$
requires a dominant NRO contribution, we can conclude from
equation~(\ref{nbase}) that in that case the sign of the second
derivative will be positive, although for large enough $k$ it will
turn to negative as the LOG part becomes dominant.

In a similar way one can obtain expressions for $P_t(k)$
[or equivalently $r(k)$] starting with the general equations (\ref{Pt}) or
(\ref{tensc}). In particular, the previous analytical approximation gives
in this case
\be
\label{tenscaap}
r(k)\simeq \frac{4 \ptwo}{\Ne}\left(1-\frac{1}{\Ne}\ln
\frac{k}{\kpiv}\right)^{-1}+
\frac{8 \ptwo A^{N+2}}{\Ne}  \left(1-\frac{\ptwo}{2N+4}\right)
\left(1-\frac{1}{\Ne}\ln \frac{k}{\kpiv}\right)^{N+1}\,.
\ee

\begin{figure}
\centering
\includegraphics[width=0.8\linewidth]{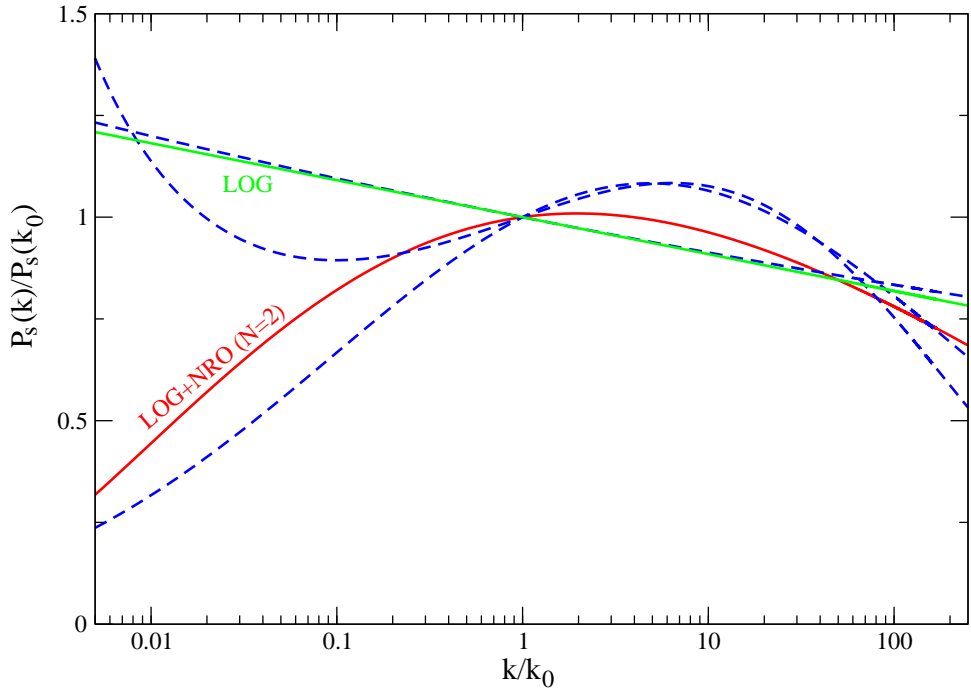}
\caption{Primordial power spectra in the standard parametrization
(dashed lines with Taylor expansion up to second order, i.e.
running of running) and as predicted by the LOG and LOG+NRO
($N=2$) scenarios. Parameters for each case are the best--fit ones
(given in Tables~\ref{tabplaw}, \ref{tabSCRGF} and
\ref{tab:SCRNRO_N2}). } \label{fig:bestfpk}
\end{figure}

Let us
finally count the number of independent parameters. From
expressions (\ref{appr22}) and (\ref{tenscaap}) we see that the
spectrum of primordial perturbations depends upon five parameters
\{$\Pks$, $\Ne$,\,$\ptwo$,\,$A$,\,$N$\}, which are combinations of
the five parameters $\{\rho_0, \beta, \phi_0, M, N\}$ appearing in
the scalar potential (\ref{VNRO}). Thus the LOG+NRO has two more
parameters than the LOG model. Again, as in the LOG case, the
smallness of $q$ implies that $P_s(k)$ is nearly independent of
$q$ and, besides, the tensor spectrum is much less important than
its scalar counterpart. Hence, $q$ will be irrelevant for a broad
range of values in the fit to the data. In practice the
primordial spectrum depends essentially on four parameters which
become just three if we consider $N$ to be a fixed integer. Again,
this is to be compared with the four parameters of the simplest
standard parametrizations (\ref{stp}) and (\ref{stp2}). In
consequence, the LOG+NRO scenario is still highly predictive.

To illustrate the shapes of the scalar power spectrum in the different
scenarios discussed in this section we plot $P_s(k)$ from (\ref{aproxx1}) in
Figure~\ref{fig:bestfpk}. The parameters for the models are chosen to be the
best--fit values given in Section~6 and discussed later in the text.

\section{Data analysis procedure}
\label{howtofit}

In order to constrain the parameters of the two
scenarios introduced above, by comparing their predictions with CMB
and LSS data, we use a modified version of the \texttt{cosmomc} package
\cite{Lewis:2002ah} with a suitable parametrization of the
expressions for $P_s(k)$ and $P_t(k)$. In the next two subsections
we describe the detailed procedure for each case separately. We
ran six chains for each model, gathering $3\times10^5$ samples per
chain, using the default Metropolis algorithm to sample the
parameter space. We discard a burn-in period encompassing the
first $10^3$ samples in each chain and we employ the usual Raftery
and Lewis mixing criterium \cite{raftery}, requiring $R-1 < 0.1$ for the 
merged
chains.

We consider a flat cosmology, taking flat priors on the
cosmological parameters $\Omega_b h^2$, $\Omega_c h^2$ (the baryon
and CDM density, respectively), $\lss$ (the ratio of the distance
to the last scattering surface to the sound horizon) and $\tau$
(the optical depth to reionization). We assume dark energy in the
form of a cosmological constant ($w=-1$) and 3 species of massless
neutrinos.  The parametrizations of the primordial power spectrum are 
discussed in detail below.

We use a combination of CMB and LSS data to constrain the
parameters of the models. The main reason for doing so is that it
helps to break parameter degeneracies \cite{Kinney:2001js}. In
particular, we use the latest WMAP \cite{Hinshaw:2006ia}, ACBAR
\cite{Kuo:2006ya}, CBI \cite{Readhead:2004gy} and BOOMERANG
\cite{Montroy:2005yx} CMB data set releases and the results from
the SDSS galaxy survey \cite{Tegmark:2006az}. We also add the
Hubble Space Telescope measurement of the Hubble constant $H_0 =
72 \pm 8$ km/s/Mpc \cite{Freedman:2000cf}.

\subsection{Logarithmic regime (LOG) }

As discussed in Section \ref{sec:paramsSCR}, the LOG scenario is
described by the three independent parameters $\,\{\rho, \beta,
\phipiv\}$, appearing in the
scalar potential (\ref{VSCR}). However, for the purpose of
comparing the model with data, it is more appropriate to use the
following set:
 \be
 \label{ParSCR}
 \mathbb{P}_{\rm LOG}\equiv\{\, \ln\Pks,
 \Ne, \ptwo\,\}\,,
 \ee
which are related to the potential parameters by the relations
(\ref{defq}), (\ref{N0}) and (\ref{Ps0}). The inverse
transformations are given by
\bea \label{SCRinversion}
\phipiv&=&\sqrt{2\ptwo\Ne}\ M_p\ , \nonumber\\
\rho_0&=&\frac{6\pi^2\ptwo \Pks}{N_e}M_p^4\ ,
\\\nonumber
\beta&=&\ptwo\frac{6\pi^2\ptwo\Pks}{N_e} M_p^4\,.
\eea

The reasons for preferring the set (\ref{ParSCR}) over the
original potential parameters are the following. First, the fit to
WMAP data is very sensitive to the value of $\ln P_s$ at the
pivotal scale, which makes it very convenient to use $\ln P_s^0$
as one of the parameters. Second, $N_e^0$ appears explicitly in
the expressions of $P_s(k)$ and $P_t(k)$ [see
equations(\ref{PsSCR}) and (\ref{PtSCR})]. In addition, $N_e^0$
has a clear physical meaning, since for small $q$ it simply
expresses the number of e-folds since the time when $k_0$ crosses
out the horizon until the end of inflation. This also makes it
possible to impose on it a physically motivated prior for the
number of e-folds, as required to solve the homogeneity and
flatness problems. Furthermore, we note from
equation~(\ref{smallrunning2}) that $N_e^0$ and the spectral index
at $k_0$ are simply related,
\bea
\label{indeapo}
n_0 \simeq 1 -\frac{1}{N_e^0} \ .
\eea
%
Finally, the third parameter, $\ptwo$, does also appear explicitly
in the expressions for the spectra. As argued in subsection
\ref{sec:paramsSCR}, we expect $\ptwo\ll 1$, implying that the
scalar primordial spectrum depends essentially only on $\{\ln
P_s^0, N_e^0\}$, while the tensor spectrum \mbox{$P_t(k)\simeq
16\epsilon P_s(k)$} is suppressed (and much less important for the
fit). Therefore it is convenient to choose $q$ as a parameter for
the fit in order to single this effect out. From the above
discussion, ${\ln P_s^0, N_e^0}$ will be well--determined by the
observable properties of the power spectrum, and therefore it is
appropriate to impose flat priors on them, which corresponds to
the assumption that they are location parameters.

The relations between $\mathbb{P}_{\rm LOG}$
and the potential parameters is non--linear and so one expects volume
effects coming from the Jacobian of the transformation that will
in general make the marginalized constraints on the potential
parameters sensitive to the choice of priors. Furthermore,
as argued above, only two combinations of parameters of the
potential are going to be well determined by the data. The constraints on
these ``principal directions'' in the potential parameter space
are however essentially prior--independent, as we discuss in
detail in Section \ref{sec:resultSCR}

Let us now focus on the physical constraints on the parameter
space spanned by $\mathbb{P}_{\rm LOG}$. The evolution equations
of the classical value of the field are based on General
Relativity. To prevent effects of quantum gravity from becoming
important, we conservatively require the energy density to satisfy
\be \label{condition2} \rho<M_p^4\,. \ee Similarly, it is sensible
to keep the inflaton field below the Planck scale. Note in
particular that, at least in this framework, the renormalization
scale $Q$ is to be identified with the value of the inflaton, in
order to maintain the radiative corrections under control, and
obviously the RGE are only reliable for $Q$ below the Planck
scale. Thus we also require, conservatively,
 \be
\label{condition1} \phipiv< M_p\,. \ee
 Notice that, since the
inflaton rolls towards zero, if the above condition is satisfied
for $\phipiv$ it will automatically be satisfied for smaller
values of $\phi$, as well. For larger values of $\phi$, imposing
equation~(\ref{condition1}) easily guarantees that they are well below
$m_p$, since there are very few e-folds before $k_0$, and they
correspond to a short range of $\phi$--values.

Moreover, one must ensure that the slow-roll approximation is
fulfilled, which means that we require
\bea
\label{epscond}
\epsilon&<&1\ , \\
|\eta|&<&1\,. \eea For simplicity we impose these conditions at
$\kpiv$ and this automatically ensures that the slow-roll is not
violated for smaller values of $k$, which means greater values of
$\phi$. Therefore the slow-roll will be guaranteed at the scale
$\kobs\equiv 10^{-4} \Mpci$, which is roughly the size of the
observable universe. On the other hand, larger values of $k$ are
probed and the slow-roll parameters grow as $\phi$ goes to zero.
Taking into account that the largest relevant multipole is about
3000, one gets a maximum $k$ around $k_{max}\equiv0.1\ \Mpci$
\cite{Leach:2002ar}. Using
equation~(\ref{phik}), we have checked that slow-roll for such
large value of $k$ is indeed satisfied by the samples in our
Markov chains. The slow-roll condition on $\eta$ is equivalent to
\be \label{ineq1} 2\Ne>1\;, \ee while the one on $\epsilon$ leads
to
 \be
 4\Ne>\ptwo.
 \ee
On the other hand, the  inequality (\ref{condition1}) implies
\be
\label{ineq2}
2\ptwo\Ne<1\,,
\ee
which together
 with
equation~(\ref{ineq1}) implies
\be
\label{conptwo} \ptwo<1\,,
\ee
as
anticipated in Subsection \ref{sec:paramsSCR}.

We found that samples in the Markov chains that fulfill the condition
(\ref{ineq2}), automatically satisfy also conditions
(\ref{ineq1}), (\ref{conptwo}) and (\ref{condition2}). This can be
understood as follows. As discussed in Subsection 4.1, we expect
a value for $\npiv\sim 0.95$, similar to the simplest fit (\ref{n1}). Then
equation~(\ref{indeapo}) implies (\ref{ineq1}).
Moreover, the value of $n$, coupled with the physical prior
(\ref{condition1}), translates into an upper bound on $q$: since
$n-1\simeq -2\eta \simeq 2 q (M_p/\phi_0)^2$, for $\phi_0\leq M_p$
one gets $q\leq (1-n)/2$, and therefore condition (\ref{conptwo})
holds. Incidentally, this upper bound
on $q$ implies that the contribution of the tensor part of the
power spectrum must be necessarily small:
$r=16\epsilon \leq 2(1-n)^2$. Finally,
(\ref{condition2}) is granted by the smallness of $\Pks$. Notice
also that the condition (\ref{epscond}) on $\epsilon$ is, in practice,
irrelevant because (\ref{conptwo}) ensures that $\epsilon<|\eta|$,
as can also be read off directly from (\ref{epsiloneta2}). In
consequence, condition (\ref{ineq2}) remains the only non-trivial
constraint.

In summary, we take $\mathbb{P}_{\rm LOG}$
[equation~(\ref{ParSCR})] as the set of independent parameters,
imposing flat priors on them and enforcing the constraint
(\ref{ineq2}). We then compute the scalar and tensor contributions
to the primordial spectrum via the expressions (\ref{PsSCR}) and
(\ref{PtSCR}).

As mentioned above, one of the reasons for choosing $N_e^0$ as an
independent parameter is its direct physical interpretation as the
number of e-folds. In fact, we have a strong theoretical prejudice
about its value, which should be\footnote{Recall
however that its value could be less than 50 if
there are subsequent episodes of inflation [see discussion after
equation~(\ref{efoldsdef})]. On the other hand, the parameter $N_e^0$ 
[defined in eq.~(\ref{N0})] could be larger
than 50 if inflation is interrupted e.g. by a waterfall condition
in hybrid models [see discussion after equation~(\ref{N0})].}$\sim50$. 
We have
taken into account this fact by performing two different analyses.
The first one imposes a flat prior on $N_e^0$, therefore assuming
no prejudice about its value and leaving the data to
constrain it. In the second case we enforce the theoretical
requirement by imposing a Gaussian prior on $N_e^0$ centered on 50
with a standard deviation of 5. The details of these two fits and
the results are discussed below, in Subsection
\ref{sec:resultSCR}.

\subsection{Logarithmic regime and non--renormalizable operator \\(LOG + NRO)}
\label{sec:paramsNRO}

Concerning the LOG+NRO scenario, for practical reasons it is
convenient to work with the set of independent parameters
 \be
\label{ParSCR2} \mathbb{P}_{\rm LOG+NRO}\equiv\{\, \ln\Pks, \Ne,
\ptwo\,, A\,, N\,\}\,,
 \ee
 where $A$ was defined in equation~(\ref{A}), instead of the parameters $\{\rho, \beta,
\phipiv, M, N\}$ of the scalar potential (\ref{VSCR}). The
relationships between $\mathbb{P}_{\rm LOG+NRO}$ and the potential
parameters are given in appendix \ref{formulae}.

\begin{figure}
\centering
\includegraphics[angle=0,width=.49\linewidth]{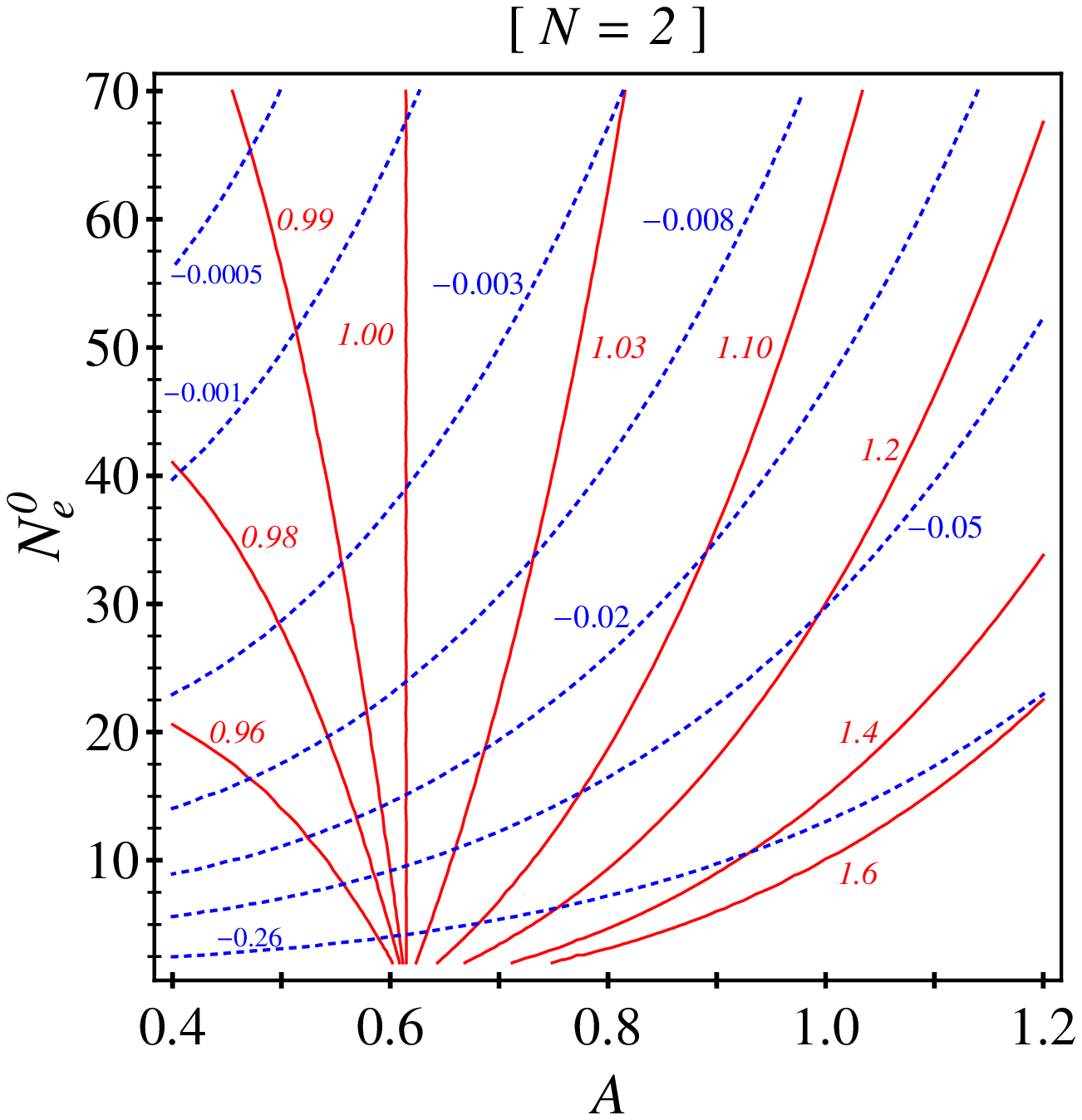}
\includegraphics[angle=0,width=.49\linewidth]{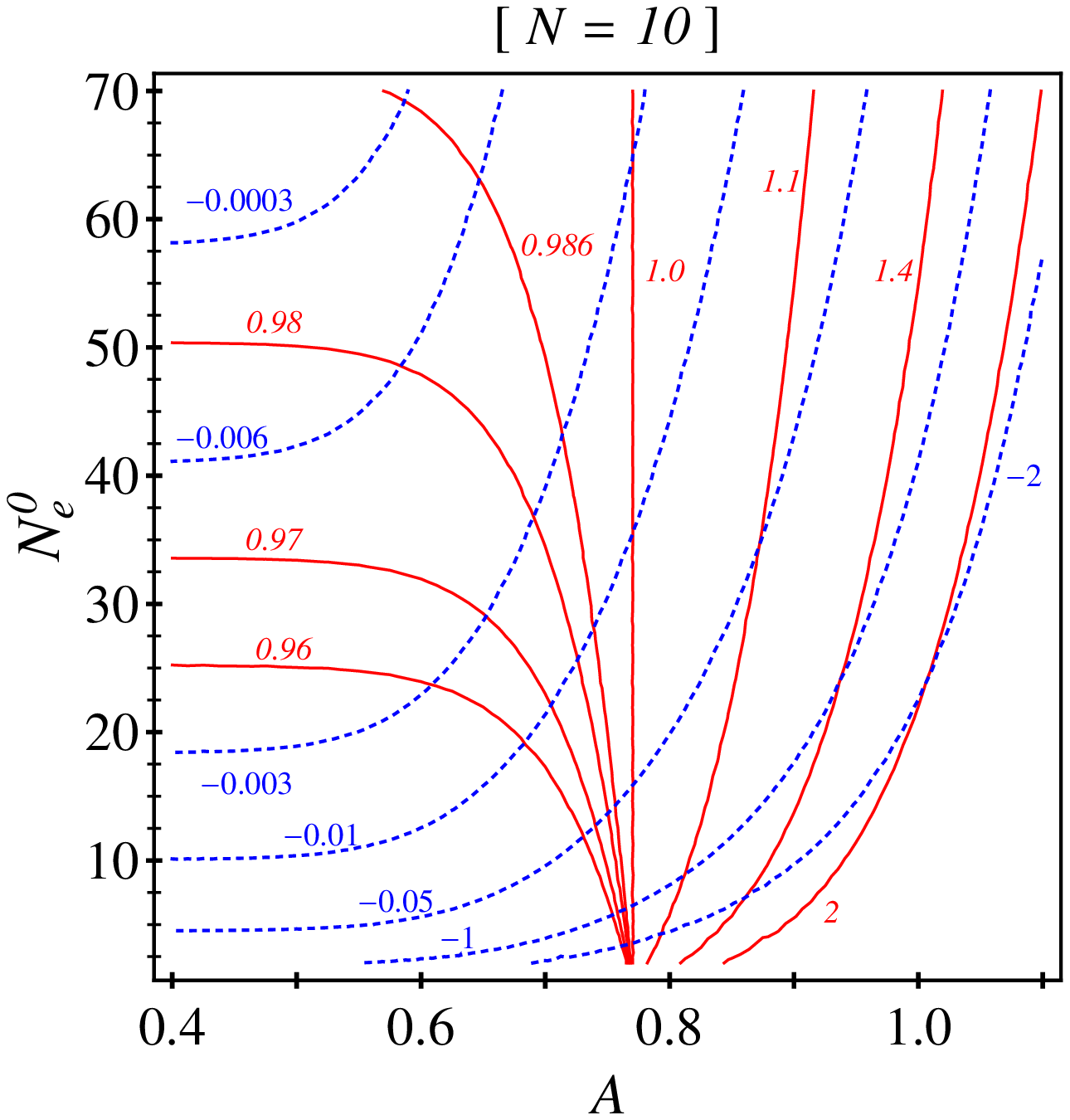}
\caption{ Values of the scalar spectral index (red lines) and its
running (blue dashed lines) at $k_0$ computed at lowest order in
slow roll for the LOG+NRO class of models with $N$ as indicated.
\label{fig:count}}
\end{figure}

The convenience and significance of the first three parameters in
$\mathbb{P}_{\rm LOG+NRO}$ are the same as in the LOG scenario.
However, the interpretation of $\Ne$ as the number of e-folds
between $k_0$ and the end of inflation is now less accurate since
there are NRO corrections, although it is still a good
approximation. This is also true for the connection between
$N_e^0$ and the spectral index $n_0$: the relation (\ref{indeapo})
becomes now \label{npivnro}
\begin{equation}
\npiv \simeq 1-\left(1+\frac{3}{2}\ptwo\right)\frac{1}{\Ne} +  \gamma
\simeq 1-\frac{1}{\Ne} +(2N+3)\frac{A^{N+2}}{\Ne}\, .
\end{equation}
This expression tells us that for not too large $N_e^0$
we should expect $A$ to be bounded from above by
some number close to unity because otherwise $\npiv$ can become
substantially different from 1 (especially for high values of $N$),
thus violating the slow-roll conditions. This is also illustrated
in Figure \ref{fig:count} which shows contour plots for the scalar
spectral index and its running at $k_0$ at lowest order in
slow-roll as functions of $A$ and $N_e^0$ for $N=2$ and $N=10$. It
is worth remarking here that since we are dealing with scale
dependent quantities the appearance of these graphs would change
if we made them at a different $k$. Figure 2 allows to realize  that, in the
context of the LOG+NRO scenario, it is possible to have simultaneously a
sizeable running and a reasonable number of e-folds.

Concerning the physical limits in parameter space, we must
take into account the presence of the scale of new physics $M$.
The role played by the Planck mass on the LOG scenario corresponds
now to $M$. To keep the validity of the effective potential
(\ref{VNRO}) the inflaton must evolve well below that scale, which
should be itself smaller than the Planck scale. Thus, we impose
the following conservative limits: \be
\label{condition4}
 2\phi_{obs}<M\leq M_p\ , \ee \be \rho<M^4\ .
\ee
 Notice that we set the first constraint at
$\phi_{obs}\equiv\phi(\kobs)$ in order to ensure it for any value
of $\phi$ in the observable range. 

As in the LOG scenario, $|\eta|\gg \epsilon$, so $\eta$ is the relevant parameter
for the breakdown of the slow-roll. However,
unlike in the LOG case, due to the NRO the absolute value of $\eta$ grows
with
 sufficiently large $\phi$. Therefore we must ensure the fulfilment of slow-roll not only at $k_{max}$ but also at $\kobs$. This guarantees that any point in between will satisfy the slow-roll conditions as well. So, we reject in the Monte Carlo process those points such that
\be |\eta(\kobs)|>\eta_{\rm lim}\ , \ee or \be
\eta(k_{max})>\eta_{\rm lim}\,, \ee being $\eta_{\rm lim}$ a
limiting value (smaller than 1) that we  set at the beginning of
the run. At the end we check that $\eta_{\rm lim}$  was indeed
well chosen to ensure the validity of the slow-roll approximation.
In practice we work with $\eta_{\rm lim}=0.2$ which is a rather
conservative value. We have checked that the change in the results is negligible
 if instead we use $\eta_{\rm lim}=0.5$.

On the other hand, we expect theoretically that the parameter $N$ should
be in the range from 1 to ${\cal O}(10)$. It is very common that flat
directions in supersymmetric models are only lifted by NROs at very high
order (as it is the case of the MSSM
\cite{Gherghetta:1995dv,Dine:1995kz}). This is also very common in D=4
string compactifications due to stringy selection rules
\cite{Cvetic:1988ez,Font:1988tp}. For
further details see \cite{Ballesteros:2005eg}. In Section \ref{results} we
discuss in detail two representative cases, which reasonably encompass the
range of values for $N$ ($N=2$ and $N=10$), and we comment on the
qualitative behaviour for values of $N$ in--between.

As for the LOG scenario, we can consider $N_e^0$ as a free
parameter with a flat prior on it or we can
alternatively constrain it to be around 50.
We have performed the two types of fit.

Finally, we can anticipate theoretically the appearance of some
strong bounds on the parameters of the model when performing the
fits. First note that the observed power spectrum normalization
$P_s^0\sim 2\times 10^{-9}$ implies through equation (\ref{Pk1})
the smallness of $\rho_0/M_p^4$. More precisely
\be
\label{rhoMpbound}
\frac{\rho_0}{M_p^4}\simeq 5\times 10^{-7}\epsilon_0 \ ,
\ee
where the subscript
``0" indicates evaluation at the pivotal scale. On the other hand,
from equations (\ref{SR_NRO}) we note that the smallness of $|\eta|$
(to preserve the slow-roll) implies that the two contributions
within the square brackets (i.e. the LOG and the NRO contributions)
must be small separately. Otherwise one should require an unjustified
fine-tuned cancellation between them. Actually, even with fine-tuning,
one could arrange the parameters to produce the cancellation
only at a particular $\phi$ (and thus $k$): since the $\phi-$dependence
of the two terms is very different, at another (not too distant)
value of $\phi$ the cancellation would not work, spoiling the slow-roll.
Consequently, the smallness of $|\eta|$ implies
 \bea
\frac{\beta}{\rho_0}\ &\lsim&
\eta_0 \left(\frac{\phi_0}{M_p}\right)^2\ ,\;\;\;
\label{etaftprev}\\
\left(\frac{\phi}{M_p}\right)^{2N+2}&\lsim&
\frac{|\eta_0|}{2(N+2)(2N+3)}\frac{\rho_0}{M_p^4}\; \simeq\;
\frac{5\times
10^{-7}\epsilon_0|\eta_0|}{2(N+2)(2N+3)}
\ ,
\label{etaft}
\eea
In the second equation we have used $M\leq M_p$ and 
equation~(\ref{rhoMpbound}). On the other hand, comparing the two
equations (\ref{SR_NRO}), and recalling that there cannot be
fine--tuned cancellations in $\eta$, it is clear that
\be
\epsilon_0\lsim \frac{1}{2} \left(\frac{\phi_0}{M_p}\right)^2
\eta_0^2\ .
\ee
Using this relation in (\ref{etaft})
we get
\be
\left(\frac{\phi}{M_p}\right)^{2N}\ \lsim\
\frac{5\times 10^{-7}|\eta_0|^3}{4(N+2)(2N+3)}\ ,
\ee
which, substituted into (\ref{etaftprev}), gives
  \be
\label{qNRO} q= \frac{\beta}{\rho_0}\lsim
 |\eta_0|^{1+\frac{3}{N}}\left[ \frac{5\times
10^{-7}}{4(N+2)(2N+3)}\right]^\frac{1}{N}\ .
 \ee
This shows that $q$ is typically small: for $N=2$ ($N=10$) one
obtains $q\lsim 1.7\times 10^{-6}$ $(q\lsim 1.4\times 10^{-2})$, a
conservative estimate obtained by replacing
$\eta_0\rightarrow\eta_{\rm lim}=0.2$. In practice $\eta$ is
substantially smaller since the bound $\eta\leq \eta_{\rm lim}$ is
to be fulfilled at all $k$, not just at the pivotal scale. Similar bounds on $q$
can be obtained using our priors together with the constraint $M\leq M_p$. 
We have checked numerically that the values obtained agree with the ones 
derived above.

\section{Results}
\label{results}

\subsection{Standard parametrization}
\label{std}

The results obtained using the standard parametrization for the primordial
spectra [equations~(\ref{stp}) and (\ref{stp2})] are summarized for easy
reference in Table \ref{tabplaw} both with and without a running spectral
index [i.e. including or not the last term of equation~(\ref{stp})]. For
later reference, we have also considered the next term in the Taylor
expansion (\ref{stp}), which has been denoted as ``running of the
running", in the last two columns of Table \ref{tabplaw}. The table shows
the best--fit parameter values, the posterior values and 68\%
1-dimensional posterior intervals for the parameters. We also give the
best--fit values for (minus twice) the log--likelihood, normalized with
respect to the model with only a constant tilt included\footnote{The
absolute value of the log--likelihood is of little interest here and in
the following. For completeness, we have computed the likelihood values
using the WMAP3 likelihood code version v2p2 with the default settings
regarding the offset for the log--likelihood. The best--fit value for the
constant tilt model is $-2\ln \mathcal{L} = 3614.0$.}. Let us recall that
the quantities describing the primordial spectrum are defined at
$\kpiv=0.002\, \textrm{Mpc}^{-1}$. These results will be useful later on
to interpret the outcomes for the LOG and LOG+NRO models and for the
comparison with them.

It is interesting to note that when using the standard
parametrization up to second order a
large and negative running of the running is preferred
\cite{Lesgourgues:2007gp}, which increases the power on large
scales (see bottom panel of Figure~\ref{fig:cls}), even though the
fit is only marginally better than the case with constant running, see 
Table \ref{tabplaw}. 
This is somewhat surprising: since in
the slow-roll approximation $n=1+2\eta-6\epsilon$, if $n$ departs
from scale--invariance too quickly then $\eta$, $\epsilon$ or both
grow up to ${\cal O}(1)$ values, marking the end of slow-roll and
the inflationary process at quite small $k$. However, in order to
solve the horizon problem we need $\sim 50-60$ e-folds of
inflation, which corresponds to the same interval in $\ln k$. This
requires the (large and negative) $d n/d \ln k$ to get suppressed
at some point, suggesting a positive second derivative (unlike the
result of the fit). This contradiction could only be avoided if
the $n(k)$ function changes abruptly at some point (or, maybe, if
there are subsequent episodes of inflation). In any case, it seems
clear that the current preference for a negative
second derivative is strongly driven by the (possibly anomalous)
low power of the large--scale multipoles. This could easily change
if observations by Planck do not confirm the lack of large scale
power observed by COBE and WMAP. However, we notice that from a
model selection perspective even present-day data do not require a
non--zero running of the running, as discussed in Section 7.
The previous discussion will be useful
later on to interpret the outcomes for the LOG and LOG+NRO models
and for the comparison with them.

\begin{table}
\caption{1-dimensional marginalized 68\% region and best fit
values results for the standard parametrizations (\ref{stp}),
(\ref{stp2}) with $n=\textrm{constant}$, $d n/\ln
k=\textrm{constant}$ and $d^2 n/\ln^2 k=\textrm{constant}$, from
left to right.} \label{tabplaw}
 \vspace{0.5cm} \centering
 \begin{tabular}{| l |l  @{\hspace{-0.06in}} r | l   @{\hspace{-0.06in}} r 
   | l  @{\hspace{-0.007in}} r |}
\hline 
&\multicolumn{2}{c |} {}  &
\multicolumn{2}{c |}{} &
\multicolumn{2}{c |}{} \\
\hspace{0.45in}Model&\multicolumn{2}{c |} {\boxed{\textrm{no running}}}  &
\multicolumn{2}{c |}{\boxed{\textrm{with running}}} &
\multicolumn{2}{c |}{\boxed{\textrm{running of running}}} \\
\hspace{0.3in}$-2 \Delta \ln \mathcal L$ &\multicolumn{2}{ c |}{$0.0$} &
\multicolumn{2}{c | }{$-3.4$} & \multicolumn{2}{c |}{$-4.4$}
\\
& 1D 68\%  & Best fit & 1D 68\%  & Best fit & 1D 68\%  & Best fit \\
\hline 
 \multicolumn{7}{|c|}{Cosmological parameters} \\\hline
 $ \Omega_b\,h^2\times10^{2} $ & $2.23 \pm 0.07$ & $2.25$ 
  & $2.20^{+0.09}_{-0.08}$ & $2.18$ & $2.22 \pm 0.08$ & $2.18$\nl
 $ \Omega_c\, h^2 $ & $0.106\pm0.004$ & $0.107$ & $0.107\pm0.004$ 
  & $0.109$ & $0.107 \pm 0.004$ & $0.107$\nl
 $ \lss$ & $1.043 \pm 0.003$ & $1.042$ & $1.043 \pm 0.003$ 
  & $1.043$ & $1.044 \pm 0.004$ & $1.043$\nl
 $ \tau$ & $0.084 \pm 0.029$ & $0.087$ & $0.114 \pm 0.035$ 
  & $0.113$ & $0.106 \pm 0.033$ & $0.109$\nl
 $H_0[{\small \textrm{Km s}^{-1} \textrm{Mpc}^{-1}}]$ & $74.3 \pm 2.1$ 
  & $73.1$ & $73.1 \pm 2.3$ & $72.0$ & $73.6 \pm 2.4$ & $72.8$  \nl \hline
 \multicolumn{7}{|c|}{Power spectra parameters} \\ \hline
 $ \ln(\Pks\times 10^{10}) $ & $3.11 \pm 0.07$ & $3.15$ 
  & $3.00 \pm 0.10$ & $3.09$ & $2.99^{+0.10}_{-0.11}$ & $3.06$\nl
 $ \npiv $ & $0.973\pm{0.019}$ & $0.961$ & $1.141^{+0.083}_{-0.082}$ 
  & $1.085$ & $1.111^{+0.096}_{-0.091} $ & $1.069$\nl
 $ \run $ & --- & --- & $-0.07 \pm 0.03$ & $-0.06$ & $-0.03 \pm 0.07$ 
  & $-0.01$\nl
 $ \runofrun $ & --- & --- & --- & --- & $-0.021 \pm 0.032$
  & $-0.032$\nl
 $ r_0 $ & $< 0.22$ & $0.003$ & $< 0.59$ & $0.15$ & $<0.63$ & $0.22$ \nl
  \hline 
\end{tabular}
\end{table}

\subsection{Logarithmic regime (LOG)}
\label{sec:resultSCR}

We denote the choice of flat prior on $N_e^0$ by LOG$\mathcal F$, where
the top--hat distribution is taken in the interval $2 \leq N_e^0 \leq
1000$. In the second case, denoted by LOG$\mathcal G$, we impose a
Gaussian prior on $N_e^0$, with mean 50 and a standard deviation of 5
e-folds.  Let us remind here that $N_e^0$ approximates the number of
e-folds since the time when the scales associated with $\kpiv$ first
crossed the horizon till the end of inflation. Thus the total number of
e-folds since the time when the largest observable scale, $\kobs$, crossed
the horizon is $\simeq N_e^0+3$. This prior choice incorporates the
theoretical prejudice that the total number of e-folds should be in the
50--60 range.

\begin{figure}
\centering
\includegraphics[angle=0,width=\linewidth]{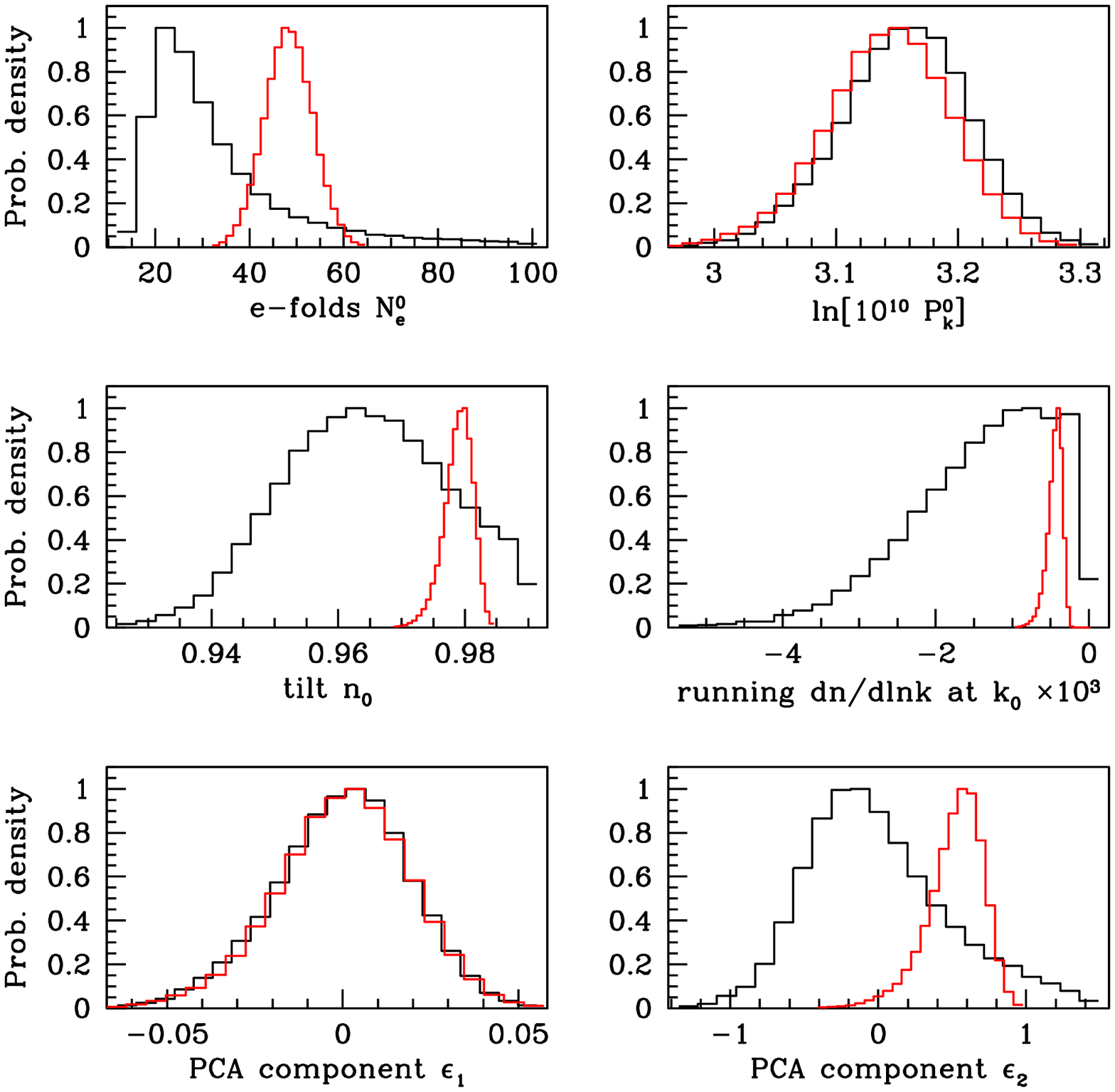}
\caption{1D marginalized probability distributions for the
well--constrained parameters in the LOG scenario (compare
Table~\ref{tabSCRGF}). Black curves are for the case with a flat
prior on $\Ne$, while red is for the case where a Gaussian prior
around $\Ne = 50$ has been enforced.
 \label{fig:SCRpars}}
\end{figure}

The  results of the Monte Carlo Markov chain (MCMC) analysis for the 
LOG$\mathcal F$ and
LOG$\mathcal G$ cases are given in Figure~\ref{fig:SCRpars} and
are summarized in Table \ref{tabSCRGF}. We give 1--dimensional
regions encompassing 68\% of probability for well--determined
parameters; robust upper bounds for parameters
whose detailed constraints are parametrization--dependent; and
best--fit values. The Table also gives posterior ranges and
best--fit values for the corresponding expressions for the tilt,
running and tensor--to--scalar ratio at the pivotal scale, $n_0$,
$dn/d\ln k|_0$ and $r_0$, respectively. These have been obtained
by using lowest order expressions in terms of the slow-roll
parameters, equations (\ref{tensc}), (\ref{n}) and (\ref{dndlnk}).

\begin{table}
\caption{Marginalized 68\% regions and best---fit values for the
class of models LOG (small--coupling regime) for quantities that
are well--determined and essentially prior/parametrization
independent. For $q$ and the tensor--to--scalar ratio at the
pivotal scale, $r_0$, we give absolute upper limits that are a
consequence of the spectral tilt and of physical priors on the
potential parameter space, Eq.~(\ref{condition1}). These bounds
have no confidence level attached as the precise numerical value
would depend on the prior/parametrization choice, a consequence
of the PCA component $\pca_3$ being an unconstrained, degenerate
direction in parameter space (see text for details).} \label{tabSCRGF} 
\vspace{0.5cm}
 \centering
 \begin{tabular}{|l | l l | l l |}
\hline 
&\multicolumn{2}{c |} {}  & \multicolumn{2}{c |}{}  \\
\hspace{0.45in}Model&\multicolumn{2}{c |}{\boxed{\textrm{LOG$\mathcal G$}} } & 
\multicolumn{2}{c |}{\boxed{\textrm{LOG$\mathcal F$}}}\\
\hspace{0.3in}$-2 \Delta \ln \mathcal L$ &\multicolumn{2}{c |}{$2.1$} & 
\multicolumn{2}{c |}{$-0.4$}\\
& 1D 68\% & Best fit & 1D 68\% & Best fit \\
\hline 
 \multicolumn{5}{| c |}{Cosmological parameters} \\\hline
 $ \Omega_b\,h^2\times10^{2} $ & $2.30\pm{0.04}$ & $2.30$ &
$2.24\pm{0.07}$ & $2.23$ \nl
  $ \Omega_c\, h^2$ &
 $0.107\pm{0.004}$ & $0.1058$ & $0.107\pm 0.004$ & $1.071$ \nl
  $ \lss$ & $1.044 \pm 0.003$ & $1.044$ &
$1.042 \pm 0.003$ & $1.041$\nl
  $ \tau$ &
$0.103 \pm 0.026$ & $0.107$ & $0.089^{+0.028}_{-0.030}$ & $0.089$
\nl 
$H_0\; [{\small \textrm{Km s}^{-1} \textrm{Mpc}^{-1}}]$ &
$74.5\pm1.5$ & $74.9$ & $73.2\pm 1.9$ & $72.9$ \nl \hline
\multicolumn{5}{|c |}{Power spectra parameters}
\\\hline
 $ \ln (\Pks\times 10^{10})$ & $3.14 \pm 0.05$ &$3.14$& $3.16^{+0.05}_{-0.06}$ & 
  $3.17$ \nl
 $ \Ne $ & $48.3 \pm 5.1$ & $48.5$ & $33.4^{+12.3}_{-12.5}$
& $25.6$ \nl
 \cline{2-5}  
 $q$ &  \multicolumn{4}{c |}{$<0.04$ for any parametrization} \nl \hline
 \multicolumn{5}{|c |}{Derived power spectra parameters}
 \\ \hline
  $ \npiv $ & $0.979\pm0.002$ & $0.979$ & $0.964^{+0.016}_{-0.013}$ & $0.961$\nl
  $ \run \times10^{3}$ & $-0.45 \pm 0.09$ & $-0.43$ & $-1.42 \pm 0.94$ &
  $-1.54$\nl
 \cline{2-5}
  $r_0$ &  \multicolumn{4}{c |}{$<4\times  10^{-3}$ for any parametrization}
  \nl \hline
 \multicolumn{5}{| c |}{Potential parameters and PCA components}
\\\hline
 $\rho/M_p^4$ &  \multicolumn{4}{c |}{$<4\times 10^{-10}$ for any parametrization} \nl
 $\beta/M_p^4$ & \multicolumn{4}{c |}{$<1\times 10^{-12}$ for any parametrization}  \nl
 $\phipiv/M_p$ & \multicolumn{4}{c |}{$<1$  (from prior, Eq.~(\ref{condition1}))}  \nl
 \cline{2-5} 
 $\pca_1$ & $0.00\pm0.02$ & 0.01 & $0.00\pm 0.02$ & 0.00 \nl
 $\pca_2$ & $0.52 \pm 0.18$ &  0.67 & $0.00\pm 0.48$ & $-0.20$  \nl
 \cline{2-5} 
 $\pca_3$ &  \multicolumn{4}{c |}{essentially unconstrained} 
  \nl \hline 
\end{tabular}
\end{table}

As anticipated, constraints on $N_e^0$ and $\ln P_s^0$ are quite tight, and 
we have checked that they are almost independent of the choice of prior by 
performing a run with priors flat in $\{\ln N_e^0, \ln q, \ln P_s^0\}$ 
instead. It is interesting to notice that in the LOG$\mathcal F$ case the 1D 
68\% (95\%) posterior region (2--tails) is approximately $21 < N_e^0 < 46$ 
($16 < N_e^0 < 81$), even though the mean is somewhat lower, at around 
33 e--folds, and a best--fit 
around 26 e--folds. This result is close to the theoretical prejudice 
$N_e^0\sim 50$, required to solve the horizon problem. Thus assuming a flat 
tree--level potential for the inflaton, the observed shape of the power 
spectrum appears to automatically point to model parameters giving a very 
sensible number of e-folds, in particular given the heavy tail of the 
probability distribution function (pdf) for large $\Ne$. This is not trivial 
at all: in principle any value for $N_e^0$ could have emerged from the 
analysis.

\begin{figure}
\centering
\includegraphics[angle=270,width=0.9\linewidth]{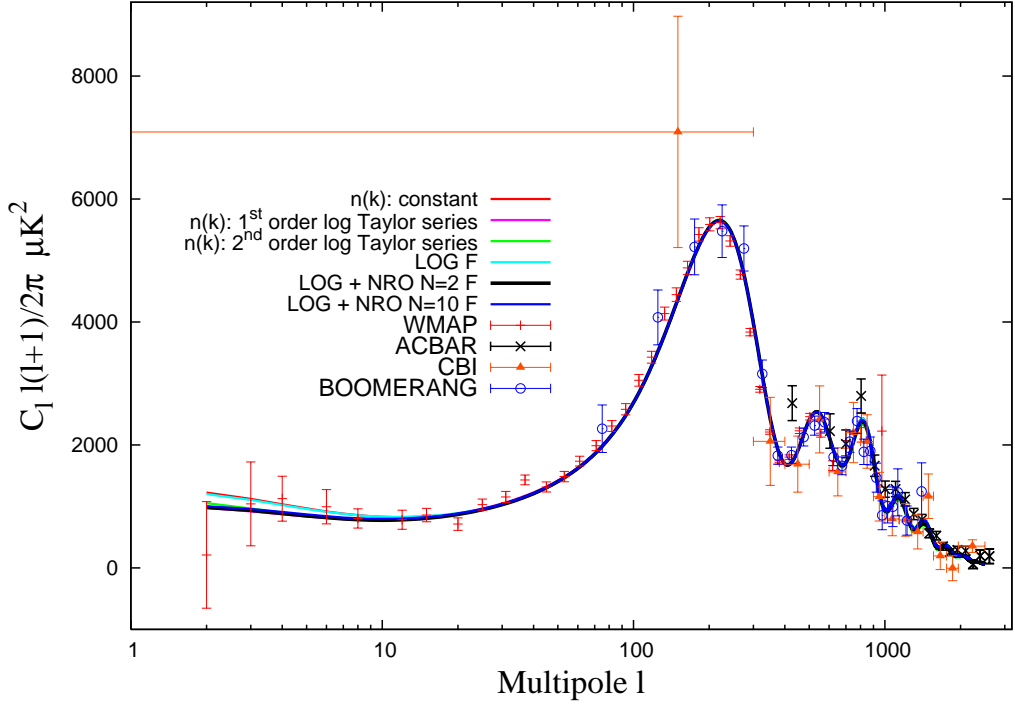}\\
\includegraphics[angle=270,width=0.9\linewidth]{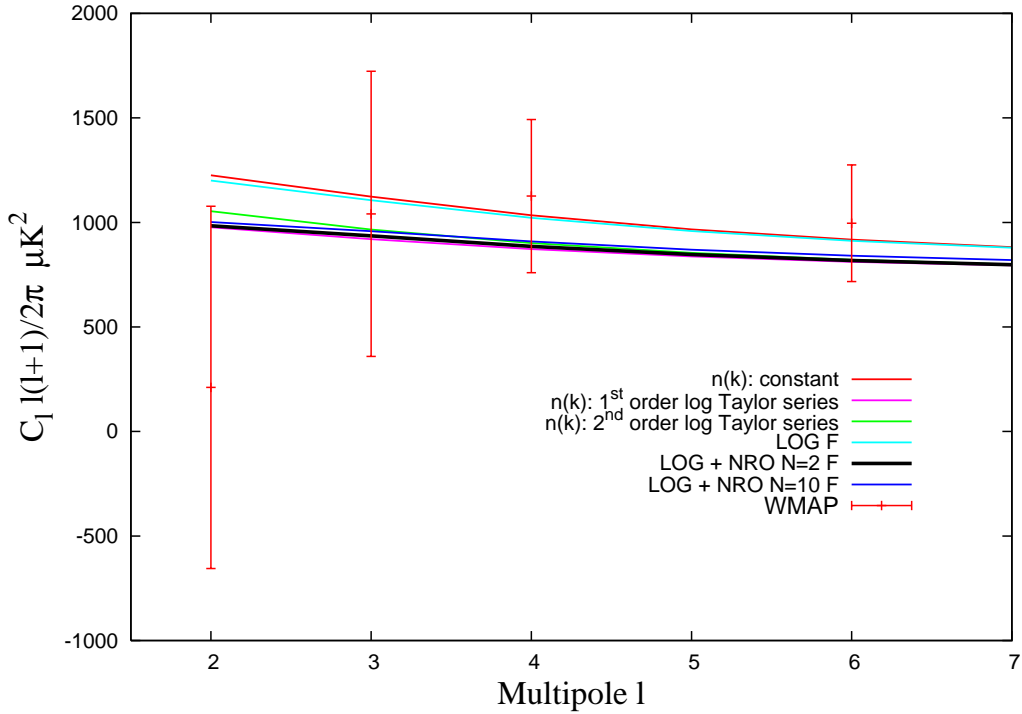}
\caption{CMB temperature power spectrum for the best--fit model
parameters for the standard parametrization, the LOG and the
LOG+NRO scenarios. The bottom panel shows the details of the
large--scale region.} \label{fig:cls}
\end{figure}

The values of the spectral index and its running at the pivotal
scale are easily derived from the parameters of the fit and are
also given in Table \ref{tabSCRGF}. Note that $dn/d\ln k|_0$ is
very small (of order $\sim 10^{-3}$), as expected from the
relation (\ref{smallrunning}). So the LOG scenario is indeed close
to the $n=$ constant limit. The value of $n$ at the pivotal scale,
$n_0$, is directly related to the value of $N_e^0$ by
equation~(\ref{indeapo}), leading to the values of $n_0$ quoted in
the Table. The LOG$\mathcal F$ best fit value, $n_0=0.961$
(corresponding to $\Ne=25.6$), coincides with the value obtained
assuming $n=$ constant and negligible running (Table 1). It is
interesting to note that although the LOG$\mathcal F$ and $n=$
constant fits are very similar, they are not identical, and indeed
LOG$\mathcal F$ gives a slightly better fit, as can be checked by
comparing the best--fit likelihood values (also compare
Figure~\ref{fig:cls}). Furthermore, if future CMB and LSS data
favour a value of $n_0$ closer to $\sim 0.98$, the value of
$N_e^0$ will come out even closer to the theoretically preferred
value, $N_e^0 \sim 50$. The upper bound on $q$ is a consequence of
the measured tilt and of the physical boundaries imposed on the
potential parameters [Recall from the discussion after
equation~(\ref{conptwo}) that we expect $q\le (1-n)/2$.] For the
reasons explained below, the pdf for $q$ depends on the prior
chosen, and therefore we do not show it in
Figure~\ref{fig:SCRpars}. However the upper bound is robust with
respect to a change of priors, and therefore we chose to report
only this value. The tensor contribution remains negligible, below
the level $\sim 10^{-3}$, since from (\ref{simpexp}) the value of
the tensor-to-scalar ratio at the pivotal scale is
 \be
 r_0 \simeq 4\frac{\ptwo}{\Ne}\;.
 \ee
Consequently the upper bound on $\ptwo$ corresponds to an order of
magnitude smaller bound on $r_0$.

The fact that we can extract two measured quantities in this
scenario (the tilt and normalization) from three model parameters
[either (\ref{ParSCR}) or (\ref{SCRinversion})] means that we
expect a strongly degenerate direction in the primordial power
spectrum parameter space. In fact, the constraints coming from the
data define a region shaped as a long solid cylinder in the 3D subspace
spanned by (\ref{SCRinversion}). Since this cylinder is not
aligned with the potential parameters direction, if one tries to
convert limits on (\ref{ParSCR}) into limits on the potential
parameters (\ref{SCRinversion}) one unavoidably picks up the
degenerate direction, i.e. along the axis of the cylinder.
This means that while in the set
(\ref{ParSCR}) the constraints on ${N_e^0, \ln P_s^0}$ are robust
with respect to a change in the parametrization of the problem (since all
the
parametrization--dependency is dumped into $q$), it is impossible
to translate these into parametrization--independent results for
the potential parameters (\ref{SCRinversion}).

However, one can still define well--constrained (and
parametrization--independent) directions in the subspace spanned
by (\ref{SCRinversion}) by performing a Principal Component
Analysis (PCA), i.e. by rotating into a new coordinate system
aligned with the degenerate direction. We therefore consider the
covariance matrix $C$ in the subspace spanned by the reduced
variables $\zeta = (\ln\hat{\rho}_0, \ln\hat{q},
\ln\hat{\phi}_0)$, where hats indicate that the variables have
been shifted by their posterior mean and normalized to their
posterior standard deviation. Then the PCA vector $\pca$ is given
by
 \begin{equation}
 \pca = U \zeta\ ,
 \end{equation}
where $U$ is the 3D rotation matrix that diagonalizes $C$:
 \begin{equation}
 \zeta^t C \zeta = \pca^t \Lambda \pca\ ,
 \end{equation}
 and $\Lambda = \text{diag}(\lambda_1, \lambda_2, \lambda_3)$ is
 the matrix of eigenvalues, whose square roots give the error along
the directions defined by $\pca$. The matrix $U$ is numerically
given by
 \begin{equation}
 U = \left(
 \begin{matrix}
 0.46 & -0.68 & 0.57 \\
 0.80 & 0.05 & -0.60 \\
 0.38 & 0.73 & 0.56
 \end{matrix}
 \right)\ ,
 \end{equation}
and $\sqrt{\lambda_1} = 0.02$, $\sqrt{\lambda_2} = 0.48$ while $
\sqrt{\lambda_3}\gg 1$, showing that $\pca_3$ is indeed the
degenerate direction. We have checked that the constraints on
$(\pca_1, \pca_2)$ are largely independent on the chosen
parametrization.

The 1-dimensional marginalized probability distributions for the
well--constrained parameters in the problem are shown in Figure
\ref{fig:SCRpars}. We do not show the probability distributions
for the non--primordial cosmological parameters as they are mostly
very similar to the standard scenario, nor
do we plot the pdf's for the parameters for which we have only
upper limits ($q,r_0, \pca_3$) since their distribution depends on the parametrization employed.

Turning now to the LOG$\mathcal G$ case, which imposes a
theoretically motivated prior on the number of e--folds, we notice
that the prior enforces $\Ne \sim 50$. This means that the model
essentially loses one further parameter, and therefore the
best--fit log--likelihood is slightly worse (see
Table~\ref{tabSCRGF}). In fact, the LOG$\mathcal G$ fit has
basically just one free parameter for the power spectrum (namely
the normalization $\ln P_s^0$), since $q$ is almost irrelevant.
Still, it gives an excellent best fit to observational data! Also,
enforcing 50 e--folds results in a very strong prediction for the
tilt to be $n_0 \sim 0.98$ (compare Figure~\ref{fig:SCRpars} and
the tightness of the posterior probability for $n_0$, red curve),
while both the tensor contribution and the running are predicted
to be very small.

In Figure~\ref{fig:cls} we plot the CMB temperature power spectrum
for the best fit models discussed here along with the compilation
of the data used.

\subsection{Logarithmic regime and non--renormalizable operator \\ (LOG +
NRO)}

In this scenario the five parameters we
use to describe the primordial spectrum are
 \be
 \mathbb{P}_{\rm LOG+NRO}\equiv\{\ln\Pks, \Ne, \ptwo\,, A\,, N\}.
 \ee
As discussed in Section \ref{sec:paramsNRO}, the meaning of the first
three ones is similar to those of the LOG case, and we impose flat priors
on $\ln\Pks$ and $\ptwo$. In analogy, we consider two types of fits:
\NROF~(with a flat prior on $N_e^0$) and \NROG~(with a Gaussian prior
centered at $N_e^0=50$ with standard deviation of $5$). Since $A$ is
expected to be of order unity or less according to the discussion above,
it is appropriate to use a flat prior on $A$ between 0 and 1.

Let us recall that $N$ determines the order of the NRO,
equation~(\ref{VNRO}). As discussed in Section
\ref{sec:paramsNRO}, we expect it to be within 1 to ${\cal
O}(10)$. One could imagine treating $N$ as a free parameter and
trying to derive a posterior bound on it from the data. However it
is technically difficult to ensure that the MCMC is correctly
performed across disjoint regions of the parameter space (since
$N$ is an integer, using it as a free parameter effectively gives
$N$ separated patches across which it is very difficult to
sample). Furthermore, NRO's with different values of $N$ are best
considered as different models, since the underlying physics is
likely to be different. Therefore distinguishing between
values of $N$ can be regarded as a model selection task, rather than a parameter
constraint exercise. For this reason it is more instructive to
consider two separate cases which are representative of the
general behaviour at low ($N=2$) and large ($N=10$) values of $N$.
Parameter constraints from CMB and LSS data, which are discussed
next, are summarized in Table~\ref{tab:SCRNRO_N2} for the $N=2$
case and in Table~\ref{tab:SCRNRO_N10} for the $N=10$ case.

\begin{table}
\caption{As in Table~\ref{tabplaw}, but for the the class of
models referred in the text as LOG+NRO, for $N=2$. Upper or lower
bounds at the specified confidence level are understood to be
1--tail limits.\label{tab:SCRNRO_N2}} \vspace{0.5cm} \centering
\begin{tabular}{|l | l l | l l |}
\hline 
&\multicolumn{2}{c |} {}  & \multicolumn{2}{c |}{}  \\
\hspace{0.45in}Model&\multicolumn{2}{c |}{\boxed{\textrm{LOG+NRO$\mathcal G$}} } & \multicolumn{2}{c |}{\boxed{\textrm{LOG+NRO$\mathcal F$}}}\\
\hspace{0.3in}$-2 \Delta \ln \mathcal L$ &\multicolumn{2}{c |}{$2.4$} & \multicolumn{2}{c |}{$-2.7$}\\
& 1D 68\% & Best fit & 1D 68\% & Best fit \\
\hline 
 \multicolumn{5}{| c |}{Cosmological parameters} \\\hline
$ \Omega_b\,h^2\times10^{2} $ & $2.32 \pm 0.05$ & $2.31$ & $2.18
\pm 0.07$ & $2.16$ \nl
  $ \Omega_c\, h^2$ & $0.107\pm{0.004}$
& $0.107$ & $0.108\pm{0.004}$ & $0.109$ \nl
  $ \lss$ & $1.044^{+0.003}_{-0.002}$ & $1.044$ & $1.041 \pm 0.003$
& $1.042$\nl
  $ \tau$ & $0.112 \pm 0.026$ & $0.010$ & $0.095\pm 0.030 $
& $0.102$\nl
  $ H_0\; [{\small \textrm{Km s}^{-1} \textrm{Mpc}^{-1}}]$ &
 $74.7^{+1.6}_{-1.5}$ & $75.0$ & $73.3^{+1.9}_{-2.0}$
& $71.6$ 
 \nl \hline
\multicolumn{5}{| c | }{Power spectra parameters}
\\\hline
$ \ln(\Pks\times 10^{10}) $ & $3.14 \pm 0.05$ & $3.13$ & $3.15 \pm
0.05$ & $3.15$\nl
 $ \Ne $ & $47.0\pm5.1$ & $49.6$ & $14.5 \pm 3.5$
& $11.7$\nl
 $ A $ & $0.46\pm0.11$ & $0.27$ &
$0.60^{+0.08}_{-0.09}$ & $0.66$ \nl
 \cline{2-5}
 $q$ &  \multicolumn{4}{c | }{$<5 \times 10^{-6}$ for any parametrization} 
 \nl \hline
 \multicolumn{5}{| c |}{Derived power spectra parameters}
 \\ \hline
  $ \npiv $ & $0.987\pm0.007$ & $0.981$ & $1.001^{+0.016}_{-0.048}$ & $1.027$\nl
  $ \run  \times 10^2 $ & $-0.11^{+0.11}_{-0.05}$ & $-0.45$ & $> -6.1\, (95\%)$ &
  $-4.28$\nl
  $\runofrun \times 10^3$ & $0.02\pm0.02$ & -0.05& $<14.2 \,(95\%)$ & $7.93$\nl
 \cline{2-5} 
  $r_0$ &  \multicolumn{4}{c |}{$<3\times  10^{-8}$ for any
  parametrization}
 \nl \hline
 \multicolumn{5}{| c | }{Potential parameters and PCA components}
 \\ \hline
 $\rho/M_p^4$ &  \multicolumn{4}{c |}{$<1 \times 10^{-10}$ for any parametrization} \nl
 $\beta/M_p^4$ & \multicolumn{4}{c |}{$<7 \times 10^{-13}$ for any parametrization}  \nl
 $\phi_0/M_p$      & \multicolumn{4}{c |}{$<1 \times 10^{-3}$  for any parametrization} \nl
 $M/M_p$      & \multicolumn{4}{c |}{$<1$  (from prior, Eq.~(\ref{condition4}))}  \nl
 \cline{2-5} 
 $\pca_1$ & $0.04\pm 0.02$ & 0.06& $0.00\pm 0.02$ & 0.00 \nl
 $\pca_2$ & $0.28 \pm 0.13$ &  0.05 & $0.01\pm 0.08$ & $-0.02$  \nl
 $\pca_3$ & $-2.3\pm 0.4$ &  -3.1 & $0.01\pm 0.72$ & $0.87$  \nl
 \cline{2-5} 
 $\pca_4$ &  \multicolumn{4}{c |}{essentially unconstrained} \nl
 \hline
\end{tabular}
\end{table}

\begin{figure}[tb]
\centering
\includegraphics[width=0.49\linewidth]{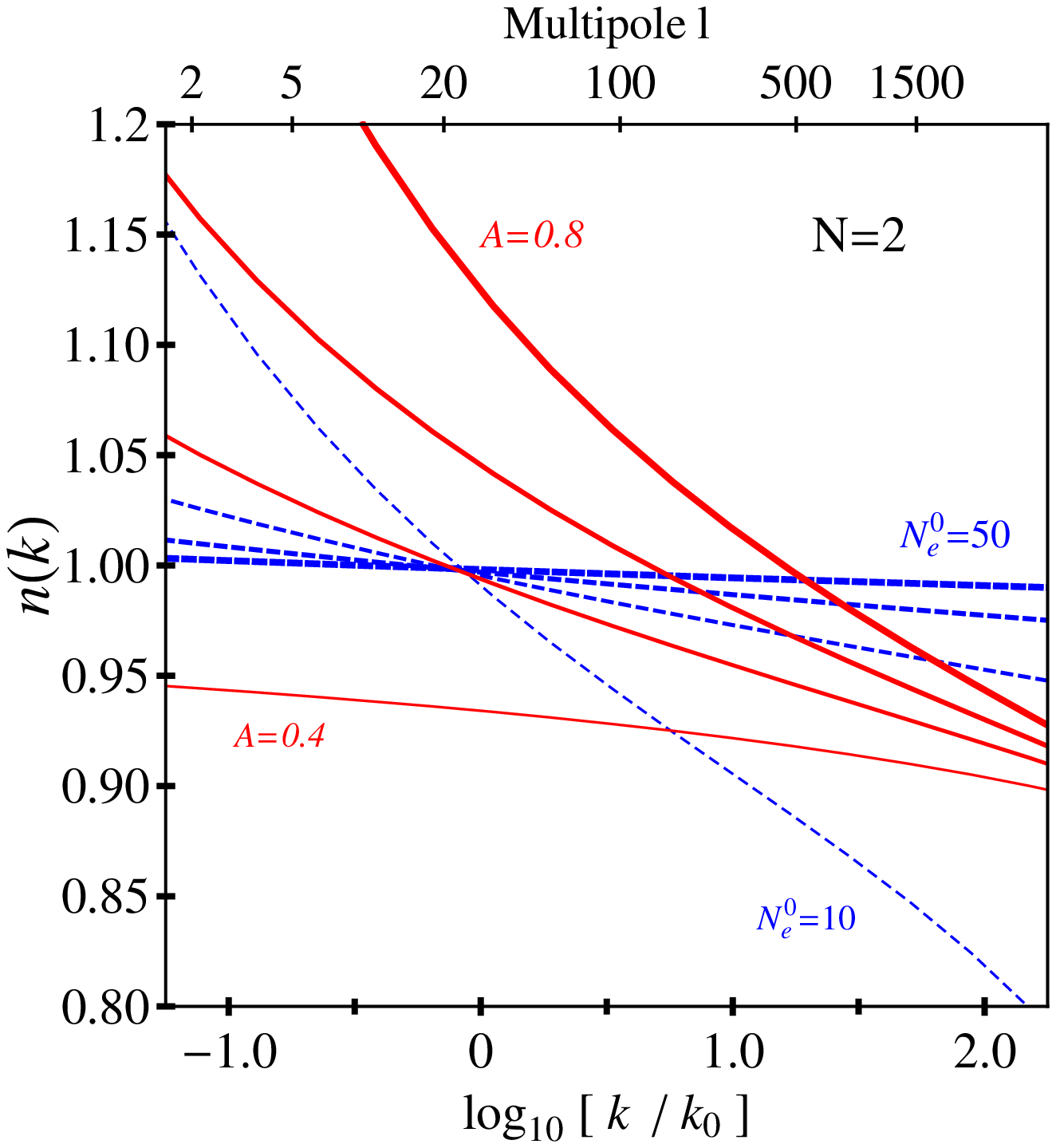}
\includegraphics[width=0.49\linewidth]{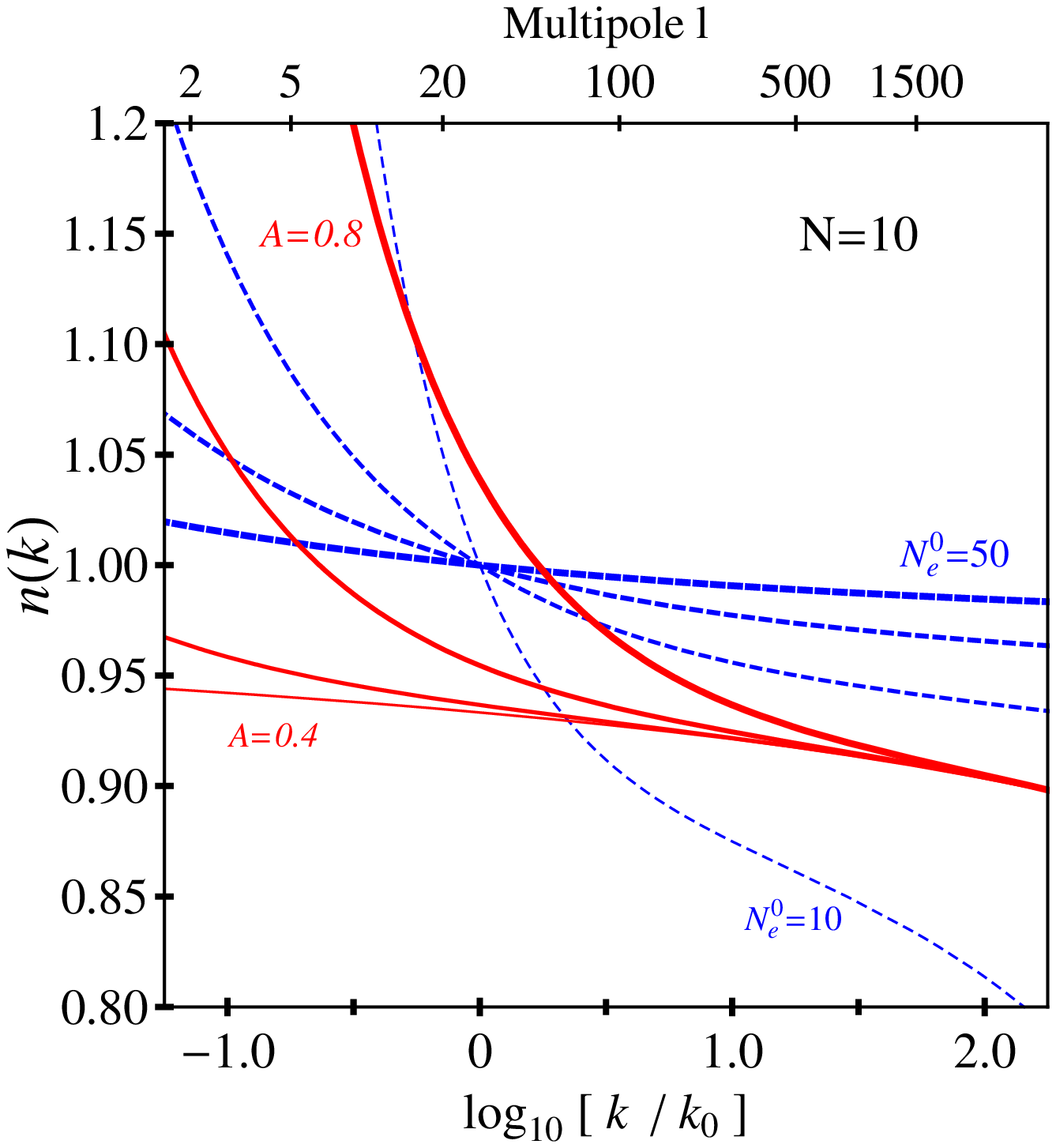}
\caption{Dependence of $n(k)$ on the parameters $\Ne$ (blue, from thin to thick
 $\Ne = 10,20,30,50$, for fixed $A=0.6$ in the $N=2$ case, left, and for 
fixed $A=0.77$ in the $N=10$ case, right) and on $A$ (red, from thin to 
thick $A = 0.2, 0.4, 0.6, 0.8$, for fixed $\Ne=15$ in both panels) in the 
LOG+NRO scenario with $N$ as indicated.} \label{fig:nkparsN2}
 \end{figure}

\begin{figure}[tb]
 \centering
 \includegraphics[width=0.49\linewidth]{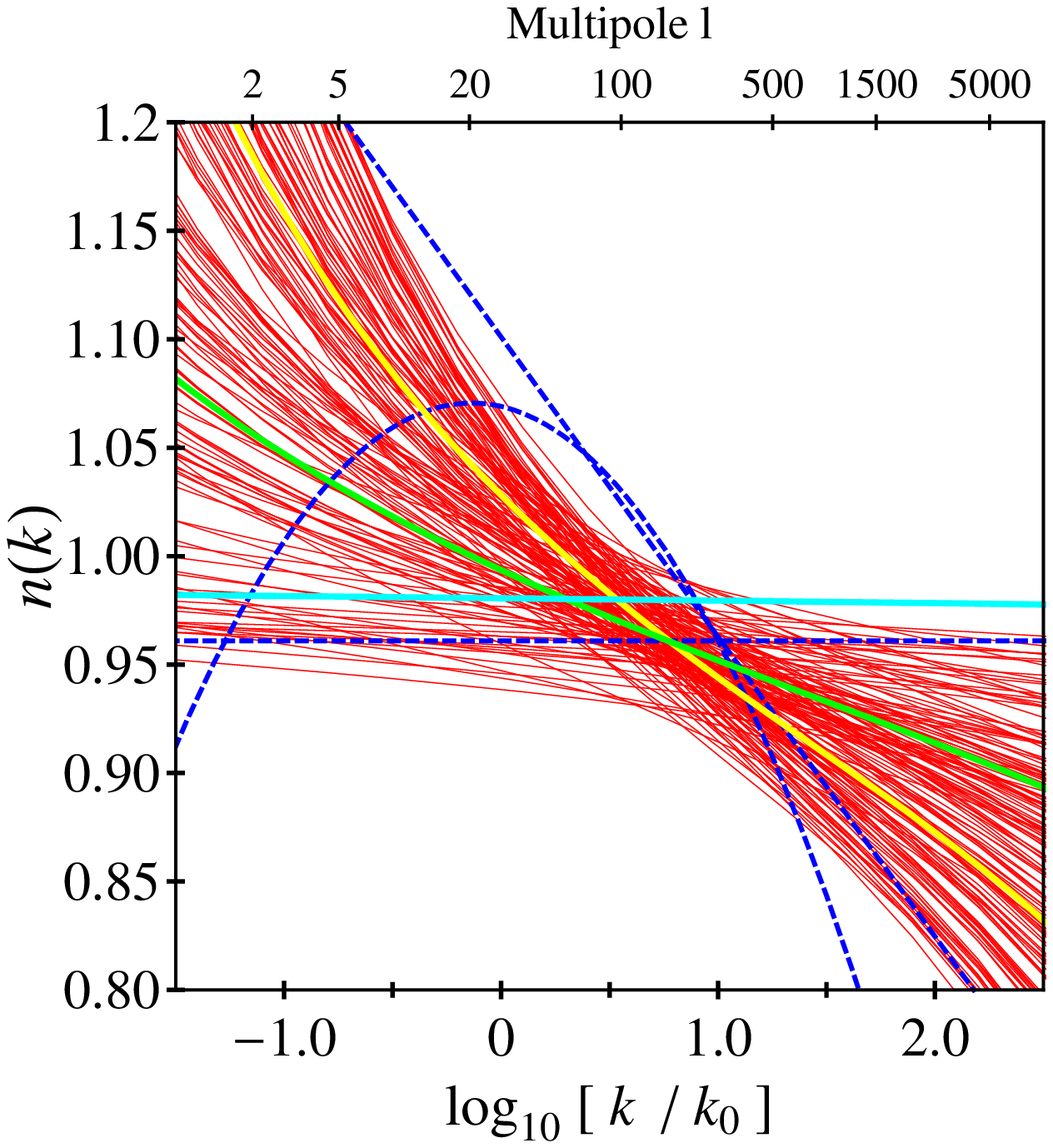}
  \includegraphics[width=0.49\linewidth]{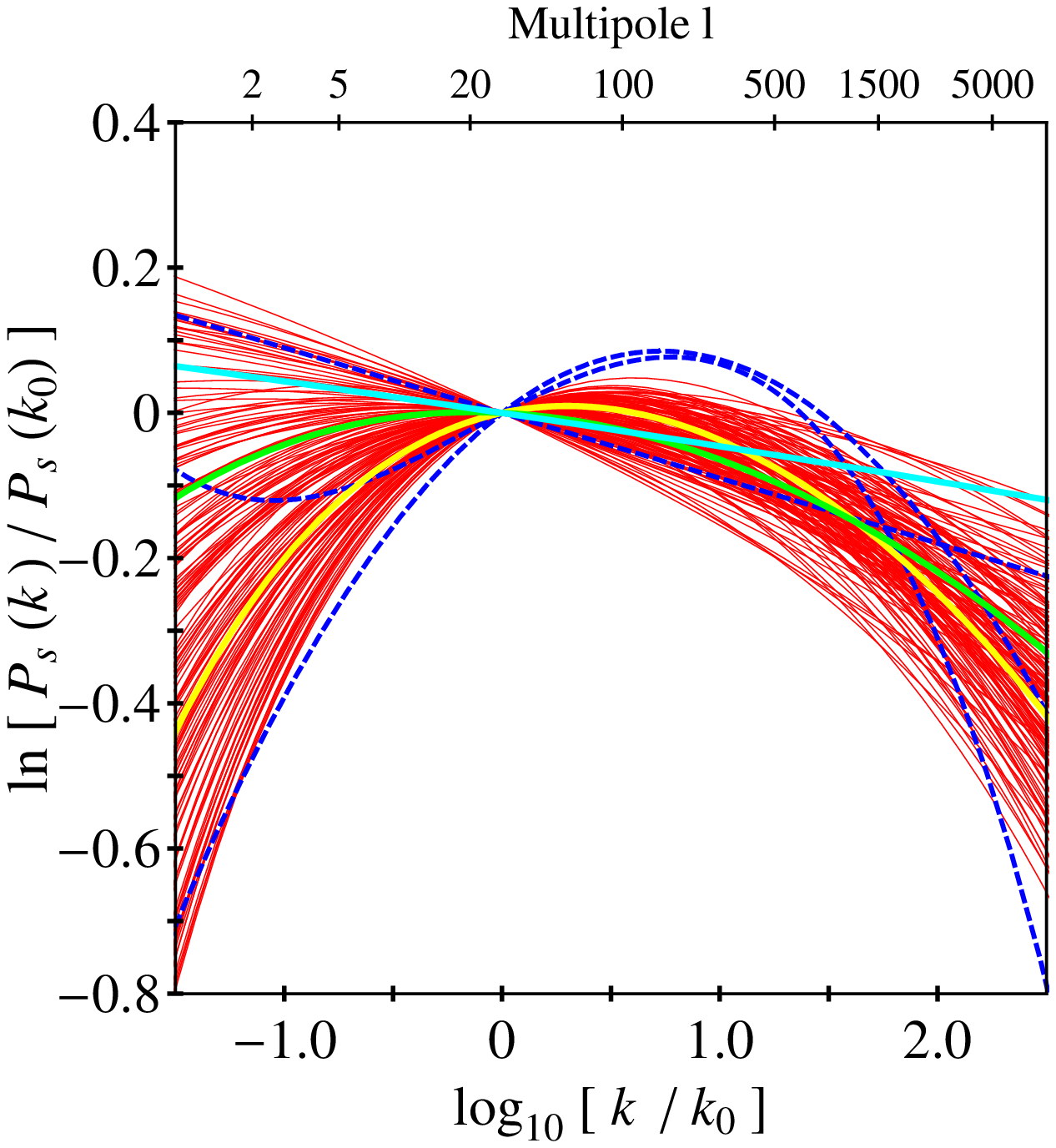}
 \caption{Preferred shape of the spectral index $n(k)$ (left) and
 the corresponding power spectrum (right) from CMB
 and LSS data (at 95\%, red curves) in the LOG+NRO scenario for
 $N=2$. The yellow line shows the best--fit
 value, the green line the posterior mean while the cyan line is the best--fit
 further imposing a Gaussian prior on the number of e--folds (\NROG~scenario).
 The dotted blue lines represent for reference
 the best--fit power spectra in the standard parametrization with tilt only,
 with running and with running of the running (from top to bottom
 on the right--hand side of the $n(k)$ panel, from top to bottom
 on the left--hand side of the $P_s(k)$ panel, compare
Table~\ref{tabplaw}).} \label{fig:nkpostN2}
 \end{figure}

Starting from the $N=2$ case, we find a strong upper bound on $q$,
which reflects the theoretical considerations exposed above and is
a consequence of the physically motivated prior
(\ref{condition4}). As a consequence, the tensor contribution is
always negligible. The number of e--folds for the \NROF~case
($N=2$) is $\Ne = 14.5 \pm 3.5$ at 68\%, becoming $10.1 \leq
\Ne \leq 25.8$ at 95\%, which is too small to solve the
horizon problem. Meanwhile, the parameter $A$ is rather tightly
constrained, $A=0.60^{+0.08}_{-0.09}$. These results can be
intuitively understood in the following way. As discussed in
Subsection~4.2, the presence of the NRO increases $d n/d \ln k$.
This effect is maximal at low $k$. The lower $N$ is, the more
gradual is the decrease of $d n/d \ln k$ with $k$. In the $N=2$
case the value of the running of the spectral index is fairly
constant in the region of $k$ accessible to observations. Thus the
model (for not too small $A$, which would lead back to the LOG
scenario) approximately resembles the $d n/d \ln k =$~constant standard
parametrization. We empirically know from the WMAP analyses
\cite{Peiris:2003ff,Spergel:2006hy} (and our own analysis in this
paper) that for this standard parametrization the value of $n$ at
$k=k_0$ cannot be very far from $n=1$. This implies from
Equations.~(\ref{nbase}), (\ref{gamma}) and the smallness of $q$ that
$A$ cannot be far from $A\sim (2N+3)^{-1/(N+2)}$, which explains
the value $A\sim 0.6$. The running $d n/d \ln k$ is then
determined by $N_e^0$ [see equation~(\ref{nbase})]. Not surprisingly,
the preferred value for the running turns out to be consistent
with the one from the standard parametrization (compare Tables
\ref{tabplaw} and \ref{tab:SCRNRO_N2}), which corresponds to the
value of $N_e^0$ quoted above. This is also consistent with our
discussion of the (too small) number of e-folds in the standard
parametrization. Some of these features are illustrated in
Figure~\ref{fig:nkparsN2} (left panel), which shows the interplay
of $\Ne$ and $A$ and their impact on the spectral index; and
Figure~\ref{fig:nkpostN2}, which shows the best--fit $n(k)$ (left
panel) and the curve corresponding to the posterior mean,
alongside with the favoured 95\% posterior region of $n(k)$ for
$N=2$. The corresponding $P_s(k)$ is shown in the right panel.

Note that the number of free parameters is essentially the same
for both the LOG+NRO case and the constant running
parametrization: 4 for the latter (3 if $r$ is set to zero) and 5
for LOG+NRO$\mathcal F$ (among which $q$ is almost irrelevant and
$N$ has been fixed), and their best--fit log--likelihoods are
similar. We comment further on this in Section~\ref{sec:modcomp}.

Further enforcing a sufficient number of e--folds by imposing a
Gaussian prior on $\Ne$ (\NROG~case in Table~\ref{tab:SCRNRO_N2})
results in a worsening of the quality of fit (an increase of minus
twice the best--fit log--likelihood by 2.1 with respect to the
standard power--law case). This is because a larger $\Ne$ for a given $A$ implies
a spectrum closer to scale--invariance, which is rather
strongly disfavoured by data (for example, \cite{Trotta:2005ar}
reports an evidence of 17:1 against a scale invariant spectrum).

As observed in the LOG case, also in the LOG+NRO case the strong
degeneracy among the potential parameters makes it impossible to
robustly translate the constraints on $\mathbb{P}_{\rm LOG+NRO}$
into prior--independent constraint for the potential parameters,
(\ref{VSCR}). As we have done above, we can still define well
constrained directions in the subspace spanned by the parameters
$\zeta = (\ln\hat{\rho}_0, \ln\hat{q}, \ln\hat{\phi}_0,
\ln\hat{M})$ (hats indicate that the variables have been shifted
by their posterior mean and normalized to their posterior standard
deviation). The eigenvalues of the 3 well--constrained directions
are now $\sqrt{\lambda_1} = 0.02$, $\sqrt{\lambda_2} = 0.08$,
$\sqrt{\lambda_3} = 0.72$ while $ \sqrt{\lambda_4}\gg 1$, and
the corresponding rotation matrix is
 \begin{equation}
 U = \left(
 \begin{matrix}
 0.49 & -0.56 & 0.43 & 0.51 \\
 -0.84 & -0.05 & 0.23 & 0.53 \\
 0.31 & 0.79 & 0.04 & 0.54 \\
 0.04 & -0.26 & -0.87 & 0.42 \\
 \end{matrix}
 \right).
 \end{equation}

\begin{figure}
\centering
\includegraphics[angle=0,width=\linewidth]{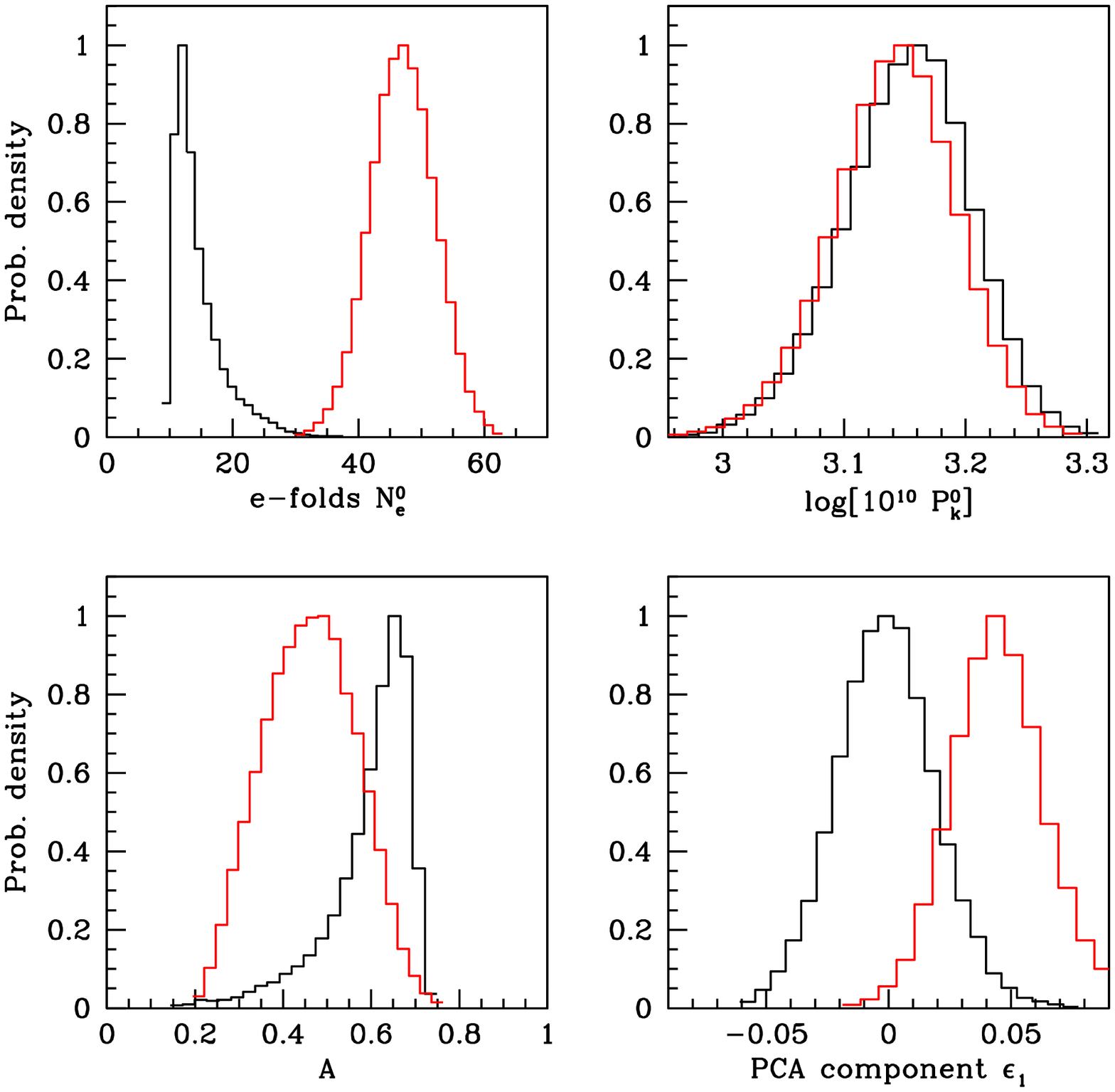}
\caption{1D marginalized probability distributions for some
well--constrained parameters in the LOG+NRO scenario, for $N=2$
(compare Table~\ref{tab:SCRNRO_N2}). Black curves are for the case
with a flat prior on $\Ne$, while red is for the case where a
Gaussian prior around $\Ne = 50$.
 \label{fig:NRO2pars}}
\end{figure}

Figure~\ref{fig:NRO2pars} shows marginalized 1--dimensional
posterior distributions for some of the well constrained
parameters in the LOG+NRO ($N=2$) scenario. Since the value of the
tilt, running and running of the running at the pivotal scale are
not really representative of the functional form of $n(k)$ in this
case, we do not show pdf's for those quantities (even though their
constraints are reported for completeness in
Table~\ref{tab:SCRNRO_N2}). A more faithful representation is
actually given in Figure~\ref{fig:nkpostN2}.

Let us now turn to the $N=10$ case. The main difference with the
previous $N=2$ case is that the effect of NRO is much more
pronounced on large scales. In particular the running $d n/d \ln
k$ can now be quite large at very low $k$ and decrease very
quickly, converging to the LOG scenario. Thus, the scenario is
qualitatively different from the constant running standard
parametrization. As for $N=2$, the value of $q$ is small and below
the theoretically expected bound given by equation~\eqref{qNRO}, which
translates again into a negligible tensor contribution ($r_0 <
3\times 10^{-4})$. The preferred value of $A$ is still
approximately determined by the empirical condition $n_0 \sim 1$.
From Equations.~(\ref{nbase}), (\ref{gamma}), this translates into
$A\sim (2N+3)^{-1/(N+2)}\sim 0.77$. The fact that the model
rapidly converges to the LOG scenario allows to increase the
number of e-folds to values similar to those of the LOG${\cal F}$
case. More precisely, in the \NROF~case, the probability
distribution for the number of e--folds has a heavier tail for
large values of $\Ne$, and the 2--tails posterior 68\% (95\%)
region is given by $18.8 < \Ne < 37.1$ ($15.8 < \Ne < 69.5$),
which appears to be solving the horizon problem within $2\sigma$.

\begin{table}
\caption{As in Table~\ref{tab:SCRNRO_N2} but for $N=10$ in the
LOG+NRO scenario. We did not perform in this case a 
Principal Component Analysis as for $N=2$. \label{tab:SCRNRO_N10} } 
\vspace{0.5cm} \centering
\begin{tabular}{|l | l l | l  l|}
\hline
&\multicolumn{2}{c |} {}  & \multicolumn{2}{c |}{}  \\
Model&\multicolumn{2}{c|}{\boxed{\textrm{LOG+NRO$\mathcal G$}} } & 
\multicolumn{2}{c|}{\boxed{\textrm{LOG+NRO$\mathcal F$}}}\\
$-2 \Delta \ln \mathcal L$ &\multicolumn{2}{c|}{$2.1$} &
\multicolumn{2}{c|}{$-1.8$}\\
& 1D 68\% & Best fit & 1D 68\% & Best fit \\
\hline
 \multicolumn{5}{| c |}{Cosmological parameters} \\\hline
$ \Omega_b\,h^2\times10^{2} $ & $2.31 \pm 0.05$ & $2.30$ & $2.23
\pm 0.07$ & $2.19$ \nl
  $ \Omega_c\, h^2$ & $0.107\pm{0.004}$
& $0.107$ & $0.107\pm{0.004}$ & $0.108$ \nl
  $ \lss$ & $1.044^{+0.003}_{-0.002}$ & $1.044$ & $1.041 \pm 0.003$
& $1.041$\nl
  $ \tau$ & $0.105 \pm 0.026$ & $0.104$ & $0.089^{+0.014}_{-0.030} $
& $0.93$\nl
  $ H_0\; [{\small \textrm{Km s}^{-1} \textrm{Mpc}^{-1}}]$ &
 $74.6\pm1.6$ & $74.5$ & $73.1\pm1.8$
& $72.1$ \nl \hline
\multicolumn{5}{| c |}{Power spectra parameters}
\\\hline
$ \ln(\Pks\times 10^{10}) $ & $3.14 \pm 0.05$ & $3.14$ & $3.15 \pm
0.05$ & $3.16$\nl
 $ \Ne $ & $48.2\pm5.1$ & $48.7$ & $28.5^{+8.6}_{-9.6}$
& $17.5$\nl
 $ A $ & $0.52^{+0.19}_{-0.16}$ & $0.54$ &
$0.57^{+0.17}_{-0.20}$ & $0.77$ \nl
 \cline{2-5}
 $q$ &  \multicolumn{4}{c |}{$<2\times 10^{-3}$ for any parametrization}
 \nl \hline
 \multicolumn{5}{| c| }{Derived power spectra parameters}
 \\ \hline
  $ \npiv $ & $0.982^{+0.002}_{-0.004}$ & $0.980$ & $0.971\pm0.016$ & $1.002$ \nl
  $ \run  \times 10^2 $ & $-0.11^{+0.09}_{-0.06}$ & $-0.05$ & $> -3.4\, (95\%)$ &
  $-4.17$\nl
  $\runofrun \times 10^3$ & $0.13^{+0.11}_{-0.18}$ & $-0.03$ & $<17.8 \,(95\%)$ & $23.7$\nl
 \cline{2-5}
  $r_0$ &  \multicolumn{4}{c |}{$<3\times  10^{-4}$ for any parametrization}
  \nl \hline
 \multicolumn{5}{| c| }{Potential parameters}
\\\hline
 $\rho/M_p^4$ &  \multicolumn{4}{c |}{$< 1 \times 10^{-11}$ for any parametrization} \nl
 $\beta/M_p^4$ & \multicolumn{4}{c |}{$<1 \times 10^{-14}$ for any parametrization}  \nl
 $\phi_0/M_p$      & \multicolumn{4}{c |}{$<0.2$  for any parametrization} \nl
 $M/M_p$      & \multicolumn{4}{c |}{$<1$  (from prior, Eq.~(\ref{condition4}))}  \nl
 \hline 
\end{tabular}
\end{table}

\begin{figure}[tb]
 \centering
 \includegraphics[width=0.49\linewidth]{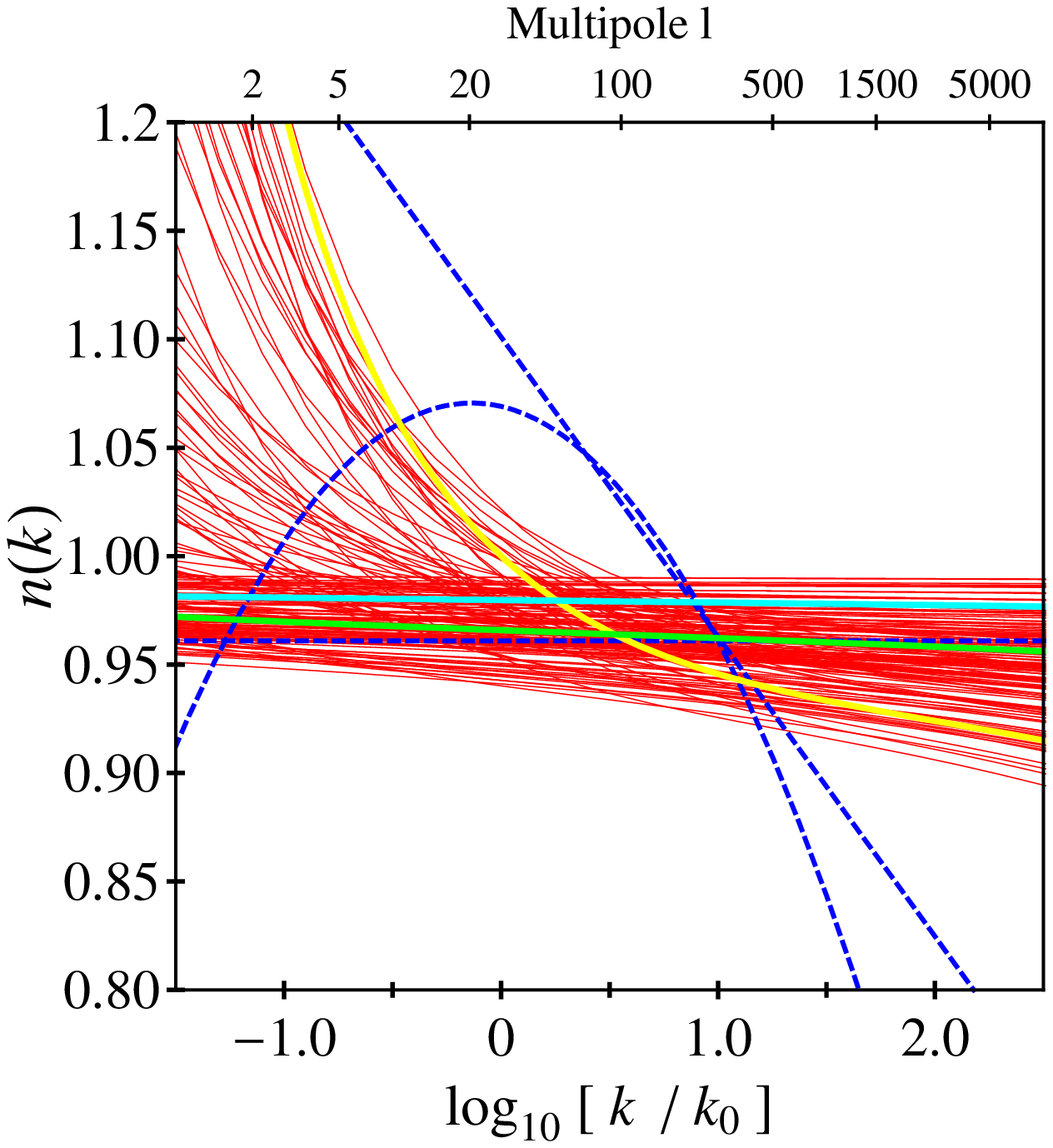}
  \includegraphics[width=0.49\linewidth]{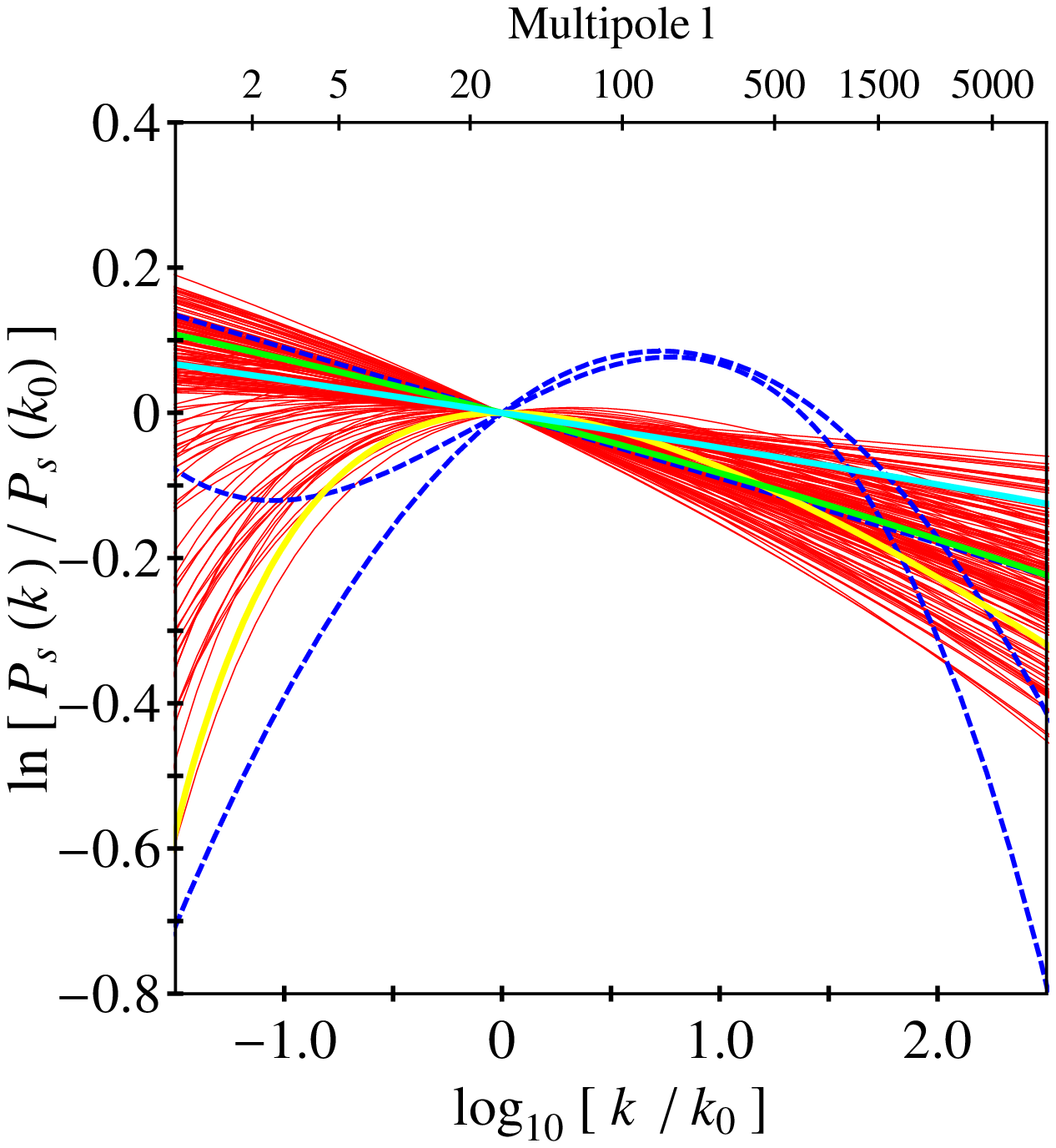}
 \caption{As in Figure~\ref{fig:nkpostN2} but for the
 LOG+NRO scenario with $N=10$.} \label{fig:nkpostN10}
 \end{figure}

Some of these features are illustrated Figure~\ref{fig:nkpostN10}.
Note in particular that the running of running for the best fit
(yellow line in Fig.~\ref{fig:nkpostN10}) is positive and
sizeable. The quality of the fit is similar to the constant
running case for the standard parametrization (though the actual
shape of the spectrum is quite different!), although slightly
worse. This is not surprising. From subsection 6.1 we know that
data prefer a negative second derivative, which cannot be achieved
in the LOG+NRO scenario at low $k$ [see discussion after
equation~(\ref{nbase})]. However, as explained there, if the spectral
index really runs, a positive second derivative can be much more
satisfactory from the physical point of view, in particular to
produce a reasonable number of e-folds. This is precisely the case
here.

If one enforces a Gaussian prior around $\Ne=50$ (\NROG~case),
then the best--fit spectrum becomes again featureless (light blue,
solid line in Figure~\ref{fig:nkpostN10}) but with a smaller tilt
than the standard parametrization ($\npiv = 0.980$ for the $N=10$
\NROG~case), which in turn means that the goodness of fit becomes
worse than the standard case (see Table~\ref{tab:SCRNRO_N10}).
Actually the model becomes in this case quite similar to the
simpler LOG{$\mathcal G$} scenario.

Finally, for values of $N$ between 2 and 10 we have found that the
behaviour is intermediate between the cases discussed in the text.

The above discussion shows that there are cases where the standard
Taylor expansion of $n(k)$ fails to capture the physics of the
models. Generally, the LOG+NRO scenario predicts a running of $n$
which is stronger on large scales. This can only be recovered with
several terms in the Taylor expansion, which results in a higher
number of free parameters in the fit. On the other hand, the
functional form of $n(k)$ in the LOG+NRO scenario implies the
positiveness of the second derivative in most of the parameter
space, unlike the standard fit. These facts make it impossible to
use the results of the standard fits to constrain the LOG+NRO
scenario: a direct comparison of the model with the data becomes
necessary (as we have done here). This situation could easily
apply to other theoretical models, as well, and therefore great
caution is necessary when interpreting generic constraints on the
coefficient of the standard Taylor expansion in terms of specific
physical models.

\section{Model comparison}
\label{sec:modcomp}

In the previous section we have presented parameter constraints
for each class of model, namely the phenomenological model in the
standard parametrization and the more physically motivated LOG and
LOG+NRO scenarios. Assessing the relative performance of the three
models is a model comparison question, to which we now turn our
attention.

In the traditional frequentist approach to statistics, model
comparison is tackled in terms of hypothesis testing: for example,
we might ask whether the improvement in the best--fit likelihood
in terms of the effective $\dchisq=-2 \Delta \ln \mathcal L$ when
adding a running to the tilt is ``significant'' enough to warrant
the inclusion of a non--zero running. There are however several
reasons why answering this question is far from trivial. A
technical reason is that the usual rule of thumb of ``$\dchisq$
per extra degree of freedom'' can only be applied if certain
regularity conditions are met, and in particular only if the extra
parameter for nested model does not lie on the boundary of the
parameter space (see \cite{Protassov:2002sz} for an astrophysical
example and references therein). So for example, the $\dchisq$
criterion could not be applied to compare the quality of fit of
the LOG model with that of LOG+NRO, since the former is obtained
from the latter by setting $A=0$, and $A<0$ is not allowed.

Another, more fundamental aspect has to do with the meaning and
interpretation of frequentist hypothesis testing. As discussed in
detail in \cite{Gordon:2007xm}, frequentist likelihood ratio tests
{\em assume} the hypothesis $\mathcal H$ is true and give the
probability of observing data $\obs$ as extreme or more extreme
than what has actually been measured. This is a statement on the
probability of the data assuming a hypothesis $\mathcal H$ to be
true (which in Bayesian terms amounts to the choice of a model,
$\mdl$), i.e. frequentist hypothesis testing gives $P(\obs |
{\mathcal H})$. But this is not the quantity one is usually
interested in, which is actually $P(\mdl|\obs)$, the probability
of the model $\mdl$ given the observations, which can only be
obtained by using Bayes theorem to invert the order of
conditioning. For this reason, model selection is an inherently
Bayesian question \cite{Trotta:2005ar}.

Bayesian model selection is based on the computation of the model
likelihood $P(\obs|\mdl) \equiv \mlike(\mdl)$ (also called
``evidence''), which is the normalization constant in the
denominator of Bayes theorem (see
\cite{Trotta:2005ar,Trotta:2007hy} for details) obtained by
averaging the likelihood $P(\obs|\theta, \mdl)$ over the prior
$P(\theta|\mdl)$ in the parameter space $\theta$ of the model

\begin{equation}
\label{eq:evidence} \mlike(\mdl) = \int P(\obs|\theta,
\mdl)P(\theta|\mdl) d\theta.
\end{equation}
From the model likelihood one obtains the model probability given
the data by using once more Bayes theorem, $P(\mdl|\obs) \propto
P(\mdl)\mlike(\mdl)$, where $P(\mdl)$ is the prior probability
assigned to the model (usually taken to be noncommittal and equal
to $1/N_m$ if one considers $N_m$ different models). When
comparing two models one usually computes the Bayes factor
$B_{12}$, given by the ratio of the evidences between the two
models:
\begin{equation}
\ln B_{12} = \ln \mlike(\mdl_1) - \ln \mlike(\mdl_2).
\end{equation}
The Bayes factor thus gives the factor by which the relative odds
among two models have changed after the arrival of the data. As a
simple calculation shows for the case of Gaussian likelihood and
prior, equation~(\ref{eq:evidence}) contains both a likelihood ratio
term which rewards better fitting, and an ``Occam's razor'' term
that disfavours unnecessary model complexity, defined in terms of
useless parameters (see \cite{Kunz:2006mc} for a discussion of
model complexity). The ``best'' model is one that combines good
fitting with model predictivity. Bayes factors are usually
interpreted against the Jeffreys' scale for the strength of
evidence, which we qualify as follows: ``weak evidence'' for $|\ln
B| < 2.5$, ``moderate evidence'' for $2.5 < |\ln B| < 5.0$ and
``strong evidence'' for $|\ln B| > 5.0$. The computation of the
model likelihood is in general a numerically difficult task, as it
involves a multi--dimensional integration over the whole of
parameter space. Furthermore, a prior dependence is (correctly)
built into the method, as the Occam's razor term depends on the
ratio of the prior to posterior volume, which gives the amount of
``wasted'' parameter space of the model. Therefore it is
problematic to evaluate the Bayes factor unless one has a
physically motivated way of setting the prior volume.

The difficulty of using a fully Bayesian approach to the
comparison of the standard Taylor series parametrization with either the LOG or
the LOG+NRO scenario is that the former represents a purely
phenomenological fit to the data, while the LOG and LOG+NRO models
are physically motivated. In particular, setting a prior on the
potential parameters of the LOG and LOG+NRO models is not
comparable to setting a strictly phenomenological prior on the
quantities of direct relevance for the fit, i.e. the spectral
tilt, the running, etc., in the standard parametrization. The
Occam's razor effect which rewards highly predictive models does
not work properly if we do not compare like with like, i.e. 
if we are unable to set priors on the parameter space of the 
phenomenological parametrizations used for the fit.
Since the standard parametrization is by
construction phenomenological, it cannot be directly compared
using the Bayesian evidence to the LOG and LOG+NRO scenarios.

However, we can still draw some interesting, partial conclusion
from a Bayesian approach. In
ref.~\cite{Gordon:2007xm}, a method was presented to derive upper
bounds on the Bayesian evidence for nested models, called
``Bayesian calibrated p--values'', that is useful in cases such as
this where there is only a very loose physical basis to assign
priors to phenomenological quantities in the fit (here, the
various terms in the expansion of the potential). This allows to
assess whether extra parameters are unnecessary within the
framework of nested models, as it gives the Bayes factor which
(under mild assumptions) {\em maximizes} the evidence in favour of
the more complex model (i.e., with more terms in the Taylor
expansion). If this turns out to be not very strong, then one can
confidently conclude that the extra parameters are not needed.

\begin{table}[!hb]
\caption{Summary of model comparison statistics. Wherever the
Bayes factor is given, the notation $\ln B_{ij}$ indicated the
Bayes factor between model $i$ and model $j$ (with $\ln B_{ij}
> 0$ favouring model $i$). An overbar indicates a prior--independent upper limit
obtained using the Bayesian calibrated p--values method. The
quantity $n$ gives the number of effective parameters in the
model. \label{tab:modcomp} } \vspace{0.5cm} \centering
\begin{tabular}{|l | l |  @{\hspace{-0.007in}} c | c |  @{\hspace{-0.007in}} r | c|}
\hline
  \hspace{0.6in} Model & $\dchisq$ & $n$ & $> 50$ e--folds? & Bayes factor & Notes
 \nl\hline
  \multicolumn{6}{| c |}{Standard parametrization}
 \\\hline
 no running & $0.0$ & $\hspace{0.05in}3(+1)$& ad hoc &  --- & \nl
 with running  & $-3.4$ & $\hspace{0.05in}4(+1)$ & ad hoc & $\ln \bar{B}_{21} =
 0.7$ & No evidence  \nl
 running of running &  $-4.4$ & $\hspace{0.05in}5(+1)$ & ad hoc & $\ln \bar{B}_{32} = 0.0$ & No evidence \nl
                        &         &   &        & $\ln
    \bar{B}_{31}= 0.4 $ & No evidence
 \nl \hline
 \multicolumn{6}{| c |}{LOG models}
 \\ \hline
 LOG$\mathcal F$ & $-0.4$ & $\sim 2$  & yes (at $2\sigma$) & --- & \nl
 LOG$\mathcal G$ & $+2.1$  & $\sim 1$ & yes  & $\ln \bar{B}_{45} = 0.4$ &
 \nl \hline
 \multicolumn{6}{| c |}{LOG+NRO models}
 \\\hline
 LOG+NRO$\mathcal F$, $N=2$ & $-2.7$ & $\sim 3$  & no & ---& Excluded \nl
 LOG+NRO$\mathcal G$, $N=2$& $+2.4$ & $\sim 2$ & yes & ---& Disfavoured \nl
 LOG+NRO$\mathcal F$, $N=10$ & $-1.8$ & $\sim 3$ & yes (at $2\sigma$) & $\ln B_{84} \sim 0.0$ & Au par LOG$\mathcal F$ \nl
 LOG+NRO$\mathcal G$, $N=10$& $+2.1$ & $\sim 2$ & yes & \hspace{0.001in} $\ln B_{95} < -2.3$ & Disfavoured 
\nl \hline
\end{tabular}
\end{table}
Table~\ref{tab:modcomp} summarizes some relevant model comparison
statistics. Focusing first on the ``standard parametrization''
section, we have employed the method of ref.~\cite{Gordon:2007xm}
to derive a prior--independent upper bound on the Bayesian
evidence in favour of extra terms in the Taylor expansion. The
maximum Bayesian evidence in favour of a running is only $\ln
\bar{B}_{21} = 0.7$ (compared to a model with just a spectral
tilt), which falls short of even the ``weak evidence'' threshold.
The maximum evidence in favour of a third term in the Taylor
expansion is even weaker. We can therefore conclude that, for the
standard parametrization, present data do not require any higher
order terms than a spectral tilt (for which
ref.~\cite{Gordon:2007xm} found a maximum evidence of $\ln \bar{B}
= 2.9$ compared to a scale--invariant spectrum). Notice that in
this phenomenological approach the number of e--folds has to be
added in by hand as an extra parameter of the model (although it
would be derivable given a specific enough model for the inflaton
potential), indicated as $(+1)$ in the column giving an
approximate value of the effective number of parameters in the
model.

Regarding the LOG class of models, the goodness of fit of the
LOG$\mathcal F$ case is similar to the one of the simple tilt
model. Although the number of free parameters of the LOG scenario
is 3, the parameter $q$ is irrelevant to the fit and therefore the
effective number of parameters is closer to 2. A more precise
counting of the effective parameters could be achieved using the
notion of Bayesian complexity~\cite{Kunz:2006mc}, but this is not
required in the context of the present discussion. It is
interesting to notice that the LOG scenario also solves the
horizon problem (within $2\sigma$ of the posterior mean) with an
extreme economy of free parameters. The LOG$\mathcal G$  case
dispenses with one further parameter (as $\Ne$ becomes almost
fixed to $\Ne \sim 50$) and the upper bound on the Bayesian
evidence in favour of LOG$\mathcal F$ indicates that the
difference of $\dchisq = 2.5$ between the flat and Gaussian prior
on $\Ne$ is not strongly significant.

The LOG model can be considered nested within the LOG+NRO
class of models, with the former formally obtained from the latter
by setting $A=0$. For $N=2$, the LOG+NRO$\mathcal F$ model falls
short of achieving the necessary number of e--folds, and for this
reason it must be excluded, even though its quality of fit is
comparable to the standard case with constant running. The
LOG+NRO$\mathcal G$ case has one extra parameter (for fixed $N$)
than LOG$\mathcal G$, and a best--fit value which is actually
slightly worse, a consequence of the Gaussian prior forcing the
posterior distribution around a value of $\Ne$ which is not
strongly favoured by the data. Hence we can conclude that
LOG+NRO$\mathcal G$ ($N=2$) is disfavoured with respect to
LOG$\mathcal G$ and LOG$\mathcal F$ since it is unable to achieve
a better fit even with one extra parameter.

The LOG+NRO$\mathcal F$ with $N=10$ has a better fit than the LOG
scenario and it achieves a sufficient number of e--folds within
$2\sigma$. The method of Bayesian calibrated p--values cannot be
used to compare the two models because the LOG model (obtained by
setting $A=0$ in the LOG+NRO model) lies at the boundary of
parameter space. However, we can still roughly estimate the Bayes
factor between LOG$\mathcal F$ and LOG+NRO$\mathcal F$ by taking a
prior width on the extra parameter $A$ of order unity (as
motivated by the theoretical expectations presented in
Section~\ref{sec:paramsNRO}) and using that, for nested models,
the Bayes factor in favour of the simpler model is approximately
(see equation~(9) in \cite{Trotta:2005ar})
 \be
 \ln B \sim I - \lambda^2/2,
 \ee
where $I$ is the logarithm of the ratio of the prior to posterior
volume (the information gain) for the extra parameter and
$\lambda$ is the number of sigmas discrepancy between the
likelihood peak and the value of the extra parameter under the
nested model (here, $A=0$). Using the values in
Table~\ref{tab:SCRNRO_N10} one obtains $\lambda \sim 2.8$ and $I
\sim 1.6$ and thus $\ln B \sim -2.3$, which would weakly favour
the LOG+NRO$\mathcal F$ model. However, one has to bear in mind
that the parameter $N$ has been fixed to a value picked among a
range of order 10 possible values --- hence one has to factor in
an extra Occam's razor effect coming from the fact that $N=10$ is
one of about 10 possible choices for $N$. Hence $\ln B$ has to be
increased by about a factor $\ln 10 = 2.3$, which brings the final
odds between LOG+NRO$\mathcal F$ and LOG$\mathcal F$ to unity
(i.e., $\ln B \sim 0$). Finally, the LOG+NRO$\mathcal G$ case has
the same quality of fit of the LOG$\mathcal G$ case and one extra
parameter. The Occam's razor term from the choice of $N$ alone
would disfavour LOG+NRO$\mathcal G$ by a factor $\ln B = 2.3$ with
respect to LOG$\mathcal G$, so even without computing the precise
Bayes factor we can conclude that this scenario is disfavoured.

In conclusion, a model comparison approach singles out the LOG
scenario and the  LOG+NRO$\mathcal F$ ($N=10$) model as the most
viable cases in light of the present data. This kind of considerations
could be extended to compare this class of models with other
inflationary scenarios, once they have been suitably parametrized
in terms of fundamental variables. However, a direct comparison
with a phenomenological approach such as the standard Taylor
expansion of the spectrum is not feasible due to the lack of
predictivity of the latter. The Bayesian evidence still concludes
that no higher--order term than the tilt is presently required in
the series.

Finally, we emphasize that the LOG and LOG+NRO models predict
tensor contributions that are generally very small and will be
largely undetectable. The most optimistic case is the LOG, where
the upper bound is of order $r_0 \sim 10^{-3}$, which might be
just within reach of future B--modes observations. Conversely, a
detection of tensor modes above $\sim10^{-3}$ would disprove the
scenario of flat tree--level inflationary potentials.

\section{Conclusions}

In this paper we have compared a broad and physically
well-motivated class of inflationary models with CMB and LSS
observational data. Namely, we have considered models with flat
tree-level potentials, which typically appear in supersymmetric
theories, where $V_{\rm tree}^{\rm SUSY}$ ordinarily has plenty of
accidental flat directions. These models, beside being very well
motivated from the physical point of view (on a similar footing to
monomial potentials), lead to very model-independent cosmological
predictions. The reason is that the potential derivatives $V'$,
$V''$,... arise from the radiative corrections to $V$, which has a
characteristic logarithmic dependence on the inflaton field. This
scenario has been labelled ``LOG" throughout the paper. In
addition, we have considered the possible presence of new physics
beyond a certain high-energy cut-off. This physics does not need
to respect the flat directions of the ``low-energy" theory, and
thus it will show up as non-renormalizable operators (NRO) in the
inflaton field, which will be dominated by the lowest--order one.
This modified scenario (labelled ``LOG+NRO") is also very well
motivated and still quite model-independent.

We have studied the performance of these scenarios when compared
with CMB and LSS. We have made first a detailed study of the
features of these models, working out both numerically accurate
results and approximate analytical expressions for $P_s(k)$,
$P_t(k)$ and other relevant quantities, such as the spectral
index, $n(k)$, as a function of suitably defined model parameters.
We also discussed the number of independent parameters and the
theoretical and phenomenological constraints on them (to be
imposed a priori). As a matter of fact, one (combination) of the
parameters is almost irrelevant, which makes these models even
more predictive. Another parameter is essentially the number of
e-folds, $\Ne$, since the time when the largest observable scales
crossed out the horizon until the end of inflation. This allows to
perform the fits in a twofold way: either leaving $\Ne$ free, and
let the data determine its value, or imposing a prior on
its value according to the usual theoretical prejudice
($\Ne=50-60$). Both approaches (labelled ${\cal F}$ and ${\cal G}$
respectively) are interesting and complementary.

In the analysis we also study the performance of standard
parametrizations of the power spectrum, based on a Taylor
expansion of $\ln P_s(k)$ and $\ln P_t(k)$ around an (arbitrary) pivotal
scale, $\ln k_0$. At first (second) order these parametrizations
correspond to a constant (constantly running) spectral index,
$n(k)=$~constant ($d n/d\ln k=$~constant). They have been used in
reference analyses, in particular by the WMAP collaboration. It is
important to keep in mind that, although useful, the standard
parametrizations are not inspired by any particularly
well-motivated inflationary physics. E.g. the results of the fit
with the $d n/d\ln k=$~constant assumption are not consistent with
a number of e-folds in the required range. Still we have also
studied them (going one order beyond the WMAP analysis) to
facilitate the discussion of the performance of the LOG and
LOG+NRO scenarios. As a general comment, care must be taken to test 
inflationary models which predict a non-negligible  scale dependence. In 
many cases the standard Taylor series 
parametrization  of equations (\ref{stp}) and (\ref{stp2}) cannot be 
accurately used in  such a situation unless a high number of terms is taken 
in the expansion.

Our main results are the following:

\begin{itemize}

\item Both the LOG and LOG+NRO scenarios predict small tensor
perturbations: $r_0\le {\cal O}(10^{-3})$.

\item The LOG scenario has essentially two parameters, $P_s(k_0)$
and $\Ne$, and implies a nearly constant $n(k)$.

\item Leaving $N_e$ as a free parameter (LOG${\cal F}$ fit), one
gets $24 < \Ne < 49$ ($16 < \Ne < 84$) at 68\% (95\%) c.l. while
the corresponding spectral index is close to $n_0=0.96$. This
result is consistent with the theoretical prejudice $\Ne\sim
50-60$, required to solve the horizon problem, which is
remarkable. In the LOG${\cal G}$ fit (i.e. imposing $\Ne=50-60$) one
gets $n_0\simeq 0.98$. Note that this fit has only one parameter
and still works very well.

\item The LOG+NRO scenario has two more
parameters, which arise from the order and the suppression scale
of the NRO. We have fixed the order of the NRO (denoted $4+2N$
throughout the paper) to two representative values ($N=2, 10$)
that reasonably encompass the sensible physical range. Thus in
practice we are playing with just one additional parameter. This
scenario can produce a sizeable running of the spectral index, and
still be consistent with the data and a reasonable number of
e-folds, especially if $N$ is not very small. The impact of the
NRO (driving a running of the spectral index) is relevant at small
$k$, corresponding to the first stages of the inflationary period,
and then it quickly converges (especially for not too small $N$)
to the LOG scenario.

\item The model comparison is delicate for several technical and
fundamental reasons. In the paper we give a fully Bayesian
discussion of the relative quality of the various scenarios
considered. Qualitatively, it can be said that the goodness of the
LOG (NRO+LOG${\cal F}$) fits is similar to the standard
$n=$~constant ($dn/d\ln k=$~constant) parametrization. The
improvement in the goodness of the fit obtained by the inclusion
of an extra parameter (as the LOG+NRO scenario implies) is not
enough (with the present data) to justify such modification, but
still it remains an interesting theoretical possibility. On the
other hand, a rigorous comparison between the evidences for LOG,
LOG+NRO scenarios and for the standard parametrizations (which are
phenomenological by construction) is not feasible. However, the
prior--independent method of the Bayesian calibrated p--values
still indicates that no higher order term than a tilt is required
in the standard Taylor expansion.

\end{itemize}

As a final conclusion, the LOG and LOG+NRO scenarios analyzed in
this paper (based on flat tree-level potentials without or with
the presence of extra physics) are not only very well motivated
from the physical point of view, but they also fit remarkably well
the CMB and LSS data, with very few parameters (the predictions
are quite model independent). In addition they are naturally
consistent with a reasonable number of e-folds. Therefore, they
can be considered as a standard physical class of inflationary
models, on a similar footing as monomial potentials.

\newpage

\newpage
\appendix

\section{Some complementary formulas}
\label{formulae}
In this appendix we provide some important expressions for the case of
the potential with NRO.
Applying the following two identities of hypergeometric functions:
\bea
_2F_1(a,b;c;z)&=& {_2F_1}(b,a;c;z)\nonumber\\
_3F_2(a,b,c;a+1,b+1;z)&=&
\frac{1}{b-a}\left[b\,{_2F_1}(a,c;a+1;z)-a\,{_2F_1}(b,c;b+1;z)\right]\nonumber
\eea
one can see, integrating equation (\ref{dlnk}), that
\begin{eqnarray}
\label{long}
\nonumber
\ln\frac{k}{\kpiv}&\simeq&
\nonumber
-\frac{\ptwo}{2}\Ne\varphi\left\{
\frac{1}{N+2}+\left(\frac{2}{\ptwo}
-\frac{1}{N+2}+\ln\varphi\right)\,{_2F_1}
\left(\frac{1}{N+2},1;\frac{N+3}{N+2};-\left(A\varphi\right)^{N+2}\right)
\right.\\
    \nonumber &+&\frac{N+2}{(N+1)^2}
\left[\left(A\varphi\right)^{-N-2}\left(1-\left(A\varphi\right)^{N+2}
\right)^{-\frac{1}{N+2}}
+\left(1-\left(A\varphi\right)^{N+2}\right)^{-\frac{1}{N+2}}\right.\\
&-&\left.\left.\left(A\varphi\right)^{-N-2}-(N+1)
{_2F_1}\left(\frac{1}{N+2},\frac{1}{N+2};\frac{N+3}{N+2};
-\left(A\varphi\right)^{N+2}\right)\right]\right\}
{\Bigg|}^{\phi^2/\phipiv^2}_{\varphi=1}
\end{eqnarray}
being the slow-roll approximation the only reason for the symbol of
approximate equality. The expression (\ref{aproxx1}) that we use in the
fits is directly obtained from (\ref{long}) neglecting the irrelevant
addends, i.e. all but the one that is not proportional to the small
parameter $\ptwo$.

The first order slow-roll parameters in this scenario are:
\bea
\label{epsilonz}
\epsilon&=&\frac{1}{4\ptwo\Ne\,\Phi\,}\left[\frac{ 1 + {\left( A\,\Phi
\right) }^{M}}
  {\frac{1}{\ptwo} +
        {\frac{1}
           {2M}{\left( A\,\Phi\right) }^{M}} +
        \frac{1}{2}\ln \Phi}\right]^2\,,\\
\label{etaz}
\eta&=&\frac{1}{2\ptwo\Ne\,\Phi\,}\,\left[\frac{(2M-1){\left( A\,\Phi
\right)
}^{M}-1}  {\frac{1}{\ptwo} +
        {\frac{1}
           {2M}{\left( A\,\Phi\right) }^{M}} +
        \frac{1}{2}\ln \Phi}\right]\,,
\eea
where we have defined
\be
\Phi\equiv\left(\frac{\phi}{\phipiv}\right)^2\ ,
\ee
to simplify the notation. Notice that in the limit of the NRO going to
zero we recover the formulas (\ref{epsiloneta2}).

The relation between $\mathbb{P}_{\rm LOG+NRO}$ and the physical
parameters of the potential are given by:
\bea
 \phipiv / M_p &=&\sqrt{2q\Ne}\ ,
\\
 \rho/M_p^4&=&48\pi^2q(N+2)^3\frac{\Pks}{\Ne}
\frac{\left(1+A^{N+2}\right)^2}{\left[2(N+2)+qA^{N+2}\right]^3}\ ,
\\
 \beta/M_p^4&=&48\pi^2q^2(N+2)^3\frac{\Pks}{\Ne}
\frac{\left(1+A^{N+2}\right)^2}{\left[2(N+2)+qA^{N+2}\right]^3}\ ,
\\
 M/M_p&=&\sqrt{2q\Ne}\left\{6\pi^2(N+2)^2\frac{\Pks}{{\Ne}^3}
\frac{A^{N+2}\left(1+A^{N+2}\right)^2}{\left[2(N+2)+qA^{N+2}\right]^3}
\right\}^{-\frac{1}{2N}}\ . \eea

\section*{Acknowledgments}

We thank C\'edric Delaunay, Massimiliano Lattanzi, Mischa
Salle and Licia Verde for interesting discussions. This work is
supported by the Spanish Ministry of Education and Science through
the research project FPA2004-02015; by a Comunidad de Madrid
project (P-ESP-00346); by a Marie Curie Fellowship of the European
Community under contract MEST-CT-2005-020238-EUROTHEPY; and by the
European Commission under
contracts MRTN-CT-2004-503369 and MRTN-CT-2006-035863 (Marie Curie
Research and Training Network ``UniverseNet''). We also
acknowledge the use of the IFT Beowulf cluster funded by the CM
project above. 
G.B. would like to thank the hospitality of
Oxford Astrophysics and Theoretical Physics for a research stay
during which this work was partly conducted. He also acknowledges
financial support from the Comunidad de Madrid and the European
Social Fund through a FPI contract.
R. RdA is supported by the program ``Juan de la Cierva'' of the Ministerio 
de
Educaci\'on y
Ciencia of Spain.
R.T. is supported by the Royal Astronomical Society
through the Sir Norman Lockyer Fellowship and by St Anne's
College, Oxford.


\begin{thebibliography}{99}
\bibitem{Peiris:2003ff}
  H.~V.~Peiris {\it et al.}  [WMAP Collaboration],
  ``First year Wilkinson Microwave Anisotropy Probe (WMAP) observations:
  Implications for inflation,''
  Astrophys.\ J.\ Suppl.\  {\bf 148} (2003) 213
  [astro-ph/0302225].
%
\bibitem{Spergel:2006hy}
  D.~N.~Spergel {\it et al.}  [WMAP Collaboration],
  ``Wilkinson Microwave Anisotropy Probe (WMAP) three year results:
  Implications for cosmology,''
  Astrophys.\ J.\ Suppl.\  {\bf 170} (2007) 377
  [astro-ph/0603449].
%
\bibitem{Guth:1980zm}
  A.~H.~Guth,
  ``The Inflationary Universe: A Possible Solution To The Horizon And 
Flatness
  Problems,''
  Phys.\ Rev.\  D {\bf 23} (1981) 347.
%
\bibitem{Mukhanov:1981xt}
  V.~F.~Mukhanov and G.~V.~Chibisov,
  ``Quantum Fluctuation And Nonsingular Universe. (In Russian),''
  JETP Lett.\  {\bf 33} (1981) 532
  [Pisma Zh.\ Eksp.\ Teor.\ Fiz.\  {\bf 33} (1981) 549].
%
\bibitem{Starobinsky:1982ee}
  A.~A.~Starobinsky,
  ``Dynamics Of Phase Transition In The New Inflationary Universe 
Scenario And
  Generation Of Perturbations,''
  Phys.\ Lett.\  B {\bf 117} (1982) 175.
%
\bibitem{Hawking:1982cz}
  S.~W.~Hawking,
  ``The Development Of Irregularities In A Single Bubble Inflationary
  Universe,''
  Phys.\ Lett.\  B {\bf 115} (1982) 295.
%
\bibitem{Guth:1982ec}
  A.~H.~Guth and S.~Y.~Pi,
  ``Fluctuations In The New Inflationary Universe,''
  Phys.\ Rev.\ Lett.\  {\bf 49} (1982) 1110.
%
\bibitem{Bardeen:1983qw}
  J.~M.~Bardeen, P.~J.~Steinhardt and M.~S.~Turner,
  ``Spontaneous Creation Of Almost Scale - Free Density Perturbations In 
An Inflationary Universe,''
  Phys.\ Rev.\  D {\bf 28} (1983) 679.
%
\bibitem{Lidsey:1995np}
  J.~E.~Lidsey, A.~R.~Liddle, E.~W.~Kolb, E.~J.~Copeland, T.~Barreiro and 
M.~Abney,
  ``Reconstructing the inflaton potential: An overview,''
  Rev.\ Mod.\ Phys.\  {\bf 69} (1997) 373
  [astro-ph/9508078].
%
\bibitem{Liddle:1994dx}
  A.~R.~Liddle, P.~Parsons and J.~D.~Barrow,
  ``Formalizing the slow roll approximation in inflation,''
  Phys.\ Rev.\  D {\bf 50} (1994) 7222
  [astro-ph/9408015].
%
\bibitem{Lyth:1998xn}
  D.~H.~Lyth and A.~Riotto,
  ``Particle physics models of inflation and the cosmological density
  perturbation,''
  Phys.\ Rept.\  {\bf 314} (1999) 1
  [hep-ph/9807278].
%
\bibitem{Copeland:1993jj}
  E.~J.~Copeland, E.~W.~Kolb, A.~R.~Liddle and J.~E.~Lidsey,
  ``Reconstructing the inflation potential, in principle and in 
practice,''
  Phys.\ Rev.\  D {\bf 48} (1993) 2529
  [hep-ph/9303288].
%
\bibitem{Gordon:2007xm}
  C.~Gordon and R.~Trotta,
  ``Bayesian Calibrated Significance Levels Applied to the Spectral Tilt 
and Hemispherical Asymmetry,''
  arXiv:0706.3014 [astro-ph].
%
\bibitem{Trotta:2005ar}
  R.~Trotta,
  ``Applications of Bayesian model selection to cosmological 
parameters,''
  Mon.\ Not.\ Roy.\ Astron.\ Soc.\  {\bf 378} (2007) 72
  [astro-ph/0504022].
%
\bibitem{legacyweb}
Nasa's Legacy Archive for Microwave Background Data
Analysis (LAMBDA), http://lambda.gsfc.nasa.gov/
%
\bibitem{Hamann:2006pf}
See however 
  J.~Hamann, S.~Hannestad, M.~S.~Sloth and Y.~Y.~Y.~Wong,
  ``How robust are inflation model and dark matter constraints from
  cosmological data?,''
  Phys.\ Rev.\  D {\bf 75} (2007) 023522
  [astro-ph/0611582].
%
\bibitem{Ballesteros:2005eg}
  G.~Ballesteros, J.~A.~Casas and J.~R.~Espinosa,
  ``Running spectral index as a probe of physics at high scales,''
  JCAP {\bf 0603} (2006) 001
  [hep-ph/0601134].
%
\bibitem{Binetruy:1996xj}
  P.~Binetruy and G.~R.~Dvali,
  ``D-term inflation,''
  Phys.\ Lett.\  B {\bf 388} (1996) 241
  [hep-ph/9606342].
%
\bibitem{handbook}
M. Abramowitz and I.A. Stegun, 
``Handbook of mathematical functions with formulas, 
graphs, and mathematical tables,'' Dover Publications, 1968.
%
\bibitem{Lewis:2002ah}
  A.~Lewis and S.~Bridle,
  ``Cosmological parameters from CMB and other data: a Monte-Carlo 
approach,''
  Phys.\ Rev.\  D {\bf 66} (2002) 103511
  [astro-ph/0205436].
%
\bibitem{raftery}
A.E. Raftery and S.M. Lewis, ``The number of iterations, convergence 
diagnostics and generic Metropolis algorithms.'' In {\em  
Practical Markov Chain Monte Carlo},
(W.R. Gilks, D.J. Spiegelhalter and S. Richardson, 
eds.), London, Chapman and Hall.
%
\bibitem{Kinney:2001js}
  W.~H.~Kinney,
  ``How to fool cosmic microwave background parameter estimation,''
  Phys.\ Rev.\  D {\bf 63}, 043001 (2001)
  [astro-ph/0005410].
%
\bibitem{Hinshaw:2006ia}
  G.~Hinshaw {\it et al.}  [WMAP Collaboration],
  ``Three-year Wilkinson Microwave Anisotropy Probe (WMAP) observations:
  Temperature analysis,''
  Astrophys.\ J.\ Suppl.\  {\bf 170}, 288 (2007)
  [astro-ph/0603451].
%
\bibitem{Kuo:2006ya}
  C.~L.~Kuo {\it et al.},
  ``Improved Measurements of the CMB Power Spectrum with ACBAR,''
  [astro-ph/0611198].
%
\bibitem{Readhead:2004gy}
  A.~C.~S.~Readhead {\it et al.},
  ``Extended Mosaic Observations with the Cosmic Background Imager,''
  Astrophys.\ J.\  {\bf 609}, 498 (2004)
  [astro-ph/0402359].
%
\bibitem{Montroy:2005yx}
  T.~E.~Montroy {\it et al.},
  ``A Measurement of the CMB  Spectrum from the 2003 Flight of 
BOOMERANG,''
  Astrophys.\ J.\  {\bf 647}, 813 (2006)
  [astro-ph/0507514].
%
\bibitem{Tegmark:2006az}
  M.~Tegmark {\it et al.},
  ``Cosmological Constraints from the SDSS Luminous Red Galaxies,''
  Phys.\ Rev.\  D {\bf 74}, 123507 (2006)
  [astro-ph/0608632].
%
\bibitem{Freedman:2000cf}
  W.~L.~Freedman {\it et al.},
  ``Final Results from the Hubble Space Telescope Key Project to Measure 
the Hubble Constant,''
  Astrophys.\ J.\  {\bf 553}, 47 (2001)
  [astro-ph/0012376].
%
\bibitem{Leach:2002ar}
  S.~M.~Leach, A.~R.~Liddle, J.~Martin and D.~J.~Schwarz,
  ``Cosmological parameter estimation and the inflationary cosmology,''
  Phys.\ Rev.\  D {\bf 66}, 023515 (2002)
  [astro-ph/0202094].
%
\bibitem{Gherghetta:1995dv}
  T.~Gherghetta, C.~F.~Kolda and S.~P.~Martin,
  ``Flat directions in the scalar potential of the supersymmetric 
standard model,''
  Nucl.\ Phys.\  B {\bf 468}, 37 (1996)
  [hep-ph/9510370].
%
\bibitem{Dine:1995kz}
  M.~Dine, L.~Randall and S.~D.~Thomas,
  ``Baryogenesis From Flat Directions Of The Supersymmetric Standard 
Model,''
  Nucl.\ Phys.\  B {\bf 458}, 291 (1996)
  [hep-ph/9507453].
%
\bibitem{Cvetic:1988ez}
  M.~Cvetic,
  ``Nonrenormalizable Sector of the Effective Lagrangian from Superstring 
Theories,'' Presented at Strings '88 Workshop, College Park, Md., Mar 
24-28, 1988. 
%
\bibitem{Font:1988tp}
  A.~Font, L.~E.~Ibanez, H.~P.~Nilles and F.~Quevedo,
  ``Degenerate Orbifolds,''
  Nucl.\ Phys.\  B {\bf 307} (1988) 109
  [Erratum-ibid.\  B {\bf 310} (1988) 764].
%
\bibitem{Lesgourgues:2007gp}
This has also been noted by
  J.~Lesgourgues and W.~Valkenburg,
  ``New constraints on the observable inflaton potential from WMAP and 
SDSS,''
  Phys.\ Rev.\  D {\bf 75}, 123519 (2007)
  [astro-ph/0703625].
%
\bibitem{Protassov:2002sz}
  R.~Protassov, D.~A.~van Dyk, A.~Connors, V.~L.~Kashyap and 
A.~Siemiginowska,
  ``Statistics: Handle with Care, Detecting Multiple Model Components 
with the
  Likelihood Ratio Test,''
  [astro-ph/0201547].
%
\bibitem{Trotta:2007hy}
  R.~Trotta,
  ``Forecasting the Bayes factor of a future observation,''
  Mon.\ Not.\ Roy.\ Astron.\ Soc.\  {\bf 378}, 819 (2007)
  [astro-ph/0703063].
%
\bibitem{Kunz:2006mc}
  M.~Kunz, R.~Trotta and D.~Parkinson,
  ``Measuring the effective complexity of cosmological models,''
  Phys.\ Rev.\  D {\bf 74}, 023503 (2006)
  [astro-ph/0602378].
%
\end{thebibliography}
\end{document}